\documentclass[10pt,preprint]{aastex}

\usepackage{amsmath}
\usepackage{mathtools}
\usepackage{epsfig}
\usepackage{graphicx}
\usepackage{tabularx}
\usepackage{multirow}
\usepackage{fixltx2e}
\usepackage{rotating}
\newcolumntype{R}{>{\raggedleft\arraybackslash}X}
\newcolumntype{C}{>{\centering\arraybackslash}X}

\def\iso#1#2{\mbox{${}^{#2}{\rm #1}$}}
\def\be1#1{\iso{Be}{1#1}}
\def\al2#1{\iso{Al}{2#1}}
\def\ca4#1{\iso{Ca}{4#1}}
\def\mn5#1{\iso{Mn}{5#1}}
\def\Fe5#1{\iso{Fe}{5#1}}
\def\fe6#1{\iso{Fe}{6#1}}
\def\zr9#1{\iso{Zr}{9#1}}
\def\tc9#1{\iso{Tc}{9#1}}
\def\pd10#1{\iso{Pd}{10#1}}
\def\sm14#1{\iso{Sm}{14#1}}
\def\hf18#1{\iso{Hf}{18#1}}
\def\pu24#1{\iso{Pu}{24#1}}

\def\pfrac#1#2{\left( \frac{#1}{#2} \right)}
\def\Ra{R_{\rm SAGB}}
\def\Rs{R_{\rm SN}}

\def\msol{M_\odot}

\def\tsn{t_{\rm travel}}
\def\tarr{t_{\rm arr}}
\def\tdep{t_{\rm dep}}
\def\Esn{E_{\rm SN}}

\def\tinit{\tarr}
\def\tfin{\tdep}

\def\tres{\Delta t_{\rm res}}
\def\dtsignal{\Delta t_{\rm signal}}

\def\beq{\begin{equation}}
\def\eeq{\end{equation}}
\def\beqar{\begin{eqnarray}}
\def\eeqar{\end{eqnarray}}

\begin{document}

\title{Astrophysical Shrapnel: \\ Discriminating Among Near-Earth Stellar Explosion Sources of Live Radioactive Isotopes}

\author{Brian J. Fry and Brian D. Fields}

\affil{Department of Astronomy, University of Illinois, Urbana, IL 61801, USA}

\author{John R. Ellis}

\affil{Theoretical Physics and Cosmology Group, Department of Physics, King's College London,
London WC2R 2LS, UK, and
Theory Division, Physics Department, CERN, CH-1211 Geneva 23, Switzerland}

\begin{abstract} 

We consider the production and deposition on Earth of isotopes with half-lives in the range 10$^{5}$ to 10$^{8}$ years that might provide signatures of nearby stellar explosions, extending previous analyses of Core-Collapse Supernovae (CCSNe) to include Electron-Capture Supernovae (ECSNe), Super-Asymptotic Giant Branch (SAGBs) stars, Thermonuclear/Type Ia Supernovae (TNSNe), and Kilonovae/Neutron Star Mergers (KNe).  We revisit previous estimates of the \fe60 and \al26 signatures, and extend these estimates to include \pu244 and \mn53.  We discuss interpretations of the \fe60 signals in terrestrial and lunar reservoirs in terms of a nearby stellar ejection $\sim 2.2$ Myr ago, showing that (i) the \fe60 yield rules out the TNSN and KN interpretations, (ii) the \fe60 signals highly constrain a SAGB interpretation but do not completely them rule out, (iii) are consistent with a CCSN origin, and (iv) are highly compatible with an ECSN interpretation.  Future measurements could resolve the radioisotope deposition over time, and we use the Sedov blast wave solution to illustrate possible time-resolved profiles.  Measuring such profiles would independently probe the blast properties including distance, and would provide additional constraints the nature of the explosion.

\begin{center}
{\tt KCL-PH-TH/2014-16}, {\tt LCTS/2014-15}, {\tt CERN-PH-TH/2014-062} 
\end{center}

\end{abstract}

\section{Introduction}

The most violent stellar explosions are the sources of most of the heavy elements on Earth, and supernovae (SNe) in particular are estimated to occur at a rate of $\sim 1-3$ per century in our Galaxy \citep[e.g.,][and references therein]{adams2013}.  It is inevitable that, over the course of geological time, some such explosions will have occurred within $\sim 100$ pc of the Earth, close enough to have deposited some ejecta on the Earth and Moon \citep[e.g.,][and references therein]{shk69,fie04}.  Indeed, the Geminga pulsar located $\sim 250$ pc away \citep{fwa2007} is the remnant of a SN explosion estimated to have occurred $\sim 300$ kyr ago, and may be partly responsible for the low density of the interstellar medium (ISM) around the Solar System \citep{bc1996}.  Similarly, \al26 gamma-ray line emission and large-angle H$\alpha$ filaments suggest a SN towards the Antlia constellation $60-240$ pc away \citep{mccu2002}; if this event created a neutron star associated with the high-proper-motion pulsar PSR J0630-2834, then the explosion occurred about 1.2 Myr ago at about 140 pc \citep{tetz2013}.  The question then arises whether some closer astrophysical explosion might have left detectable traces on the Earth itself in the form of geological isotope anomalies.  Moreover, with a closer astrophysical explosion, the possibility for biological damage, even a mass extinction arises \citep[for recent references, see, e.g.,][]{melt11, beech11, dart11, atri14}.

Discussions of this possibility date back to the pioneering study of \citet{alva80}.  These authors discovered an iridium anomaly associated with the Cretaceous-Tertiary transition that they argued could not, in fact, be associated with a SN explosion, but instead with a giant impact.  Subsequently, \citet{efs96} surveyed possible isotope signatures of a nearby SN explosion, including \al26, \mn53, \fe60, and \pu244.  Motivated by this study, \citet{knie99} searched for an anomaly in the \fe60 abundance in a deep-ocean ferro-manganese (Fe-Mn) crust, and found one that appeared $\sim 2.2$ Myr ago.  Although primordial Solar System composition shows enrichment from extra-solar origins, to our knowledge, this is the first such specific extra-solar event to be identified.  Following the \citet{knie99} discovery, its interpretation was discussed in \citet{fe99} and possible corroborating isotope signatures were discussed in \citet{fhe05}.  \citet{bene2002} proposed that the event arose in the Sco-Cen OB association, which was $\sim 130$ pc away at the time of the \fe60-producing event.  \citet{faj08} presented hydrodynamic models for the SN blast impact with the solar wind, and \citet{af11} highlighted the importance of the ejecta condensation into dust grains.  

The \fe60 signal has subsequently been confirmed in another Fe-Mn crust sample \citep{knie04,fit08} and in lunar regolith \citep{cook09,fimi12,fimi14}, but no other accompanying isotope anomaly has been found in studies of \al26 abundances \citep{feige13}. Searches for \pu244 have produced just a single count, albeit with no stable isobar background \citep{wall00,wall04}.

In this paper, we broaden our previous analyses in four ways.  In a first step, we provide yields for isotopes from the CCSNe considered previously and extend our analysis to include the cases of ECSNe, TNSNe (also known as a Type Ia SN), KNe (also known as Neutron Star Mergers), and SAGBs, which have not been considered previously in this context.  For this paper, we distinguish between ECSNe and the more massive CCSNe since there are qualitative differences in the collapse and explosion mechanism as well as nucleosynthesis of these two classes.  Secondly, we revisit the formalism surrounding the deposition calculations, including the impact and some geology of the uptake factor, and also the possibility of using sediments to get time-resolved signals and give predictions for these profiles.  We also discuss the filtering processes impacting the transport of the signal via dust.  Next, we discuss the compatibility between the terrestrial and lunar evidence for a \fe60 anomaly, and we also analyze the existing limits on the \al26 abundance from samples bracketing the \fe60 anomaly.  In combination with these previous steps, as a fourth and final step, we survey the possible interpretations of the \fe60 anomaly and make predictions for upcoming measurements.

We find that a TNSN and a KN would yield too little \fe60, and can be ruled out as possible sources for the \citet{knie04} \fe60 signal.  Additionally, we find a SAGB source constrained, but not eliminated due to uncertainty in the Local Bubble's magnetic field and the location of a possible SAGB source.  CCSNe from our set of masses and ECSNe can not be ruled out based on the available measurements.

\section{Progenitors and Delivery to the Solar System}
\label{sect:progenitors}

Previous papers have focused on CCSNe as the likeliest progenitor for the \fe60 signal.  However, there are other astrophysical ejections that are thought to produce \fe60 but have not been considered previously.  These include TNSNe, ECSNe, KNe, and SAGBs.  Table \ref{tab:yields} summarizes the yields for possible CCSNe, ECSNe, and SAGBs progenitors as used in our model calculations; yields are expressed in units of $\msol$.

\begin{center}
\begin{table}[*h]
\caption{Ejected Masses for Various Radioactive Isotopes, in $\msol$
\label{tab:yields}
}
	\begin{tabularx}{\textwidth}{ |X|C|C|C|C|C|C| } \hline \hline
			Progenitor	&	15-$\msol$ CCSN$^{a}$	&	19-$\msol$ CCSN$^{a}$	&	20-$\msol$ CCSN$^{a}$	&	21-$\msol$ CCSN$^{a}$	&	25-$\msol$ CCSN$^{a}$	&	8-10-$\msol$ ECSN$^{b}$	\\  \hline
			\al26	&	$2.6 \times 10^{-5}$		&	$3.2 \times 10^{-5}$		&	$3.0 \times 10^{-5}$		&	$4.6 \times 10^{-5}$		&	$7.0 \times 10^{-5}$		&	$4.4 \times 10^{-8}$		\\  \hline
			\mn53	&	$1.8 \times 10^{-4}$		&	$2.1 \times 10^{-4}$		&	$1.3 \times 10^{-4}$		&	$2.3 \times 10^{-4}$		&	$3.6 \times 10^{-4}$		&	$1.1 \times 10^{-6}$		\\  \hline
			\fe60	&	$6.6 \times 10^{-5}$		&	$1.1 \times 10^{-4}$		&	$3.6 \times 10^{-5}$		&	$2.5 \times 10^{-5}$		&	$1.5 \times 10^{-4}$		&	$3.6 \times 10^{-5}$		\\  \hline
			\ca41	&	$4.3 \times 10^{-6}$		&	$2.7 \times 10^{-5}$		&	$4.3 \times 10^{-4}$		&	$6.9 \times 10^{-6}$		&	$3.2 \times 10^{-5}$		&	$2.0 \times 10^{-7}$		\\  \hline
			\zr93	&	$1.3 \times 10^{-8}$		&	$4.7 \times 10^{-8}$		&	$9.8 \times 10^{-9}$		&	$5.9 \times 10^{-8}$		&	$1.5 \times 10^{-7}$		&	N/A$^{d}$					\\  \hline
			\tc97	&	$4.8 \times 10^{-11}$		&	$4.2 \times 10^{-11}$		&	$1.9 \times 10^{-10}$		&	$1.3 \times 10^{-10}$		&	$8.3 \times 10^{-11}$		&	N/A$^{d}$					\\  \hline
			\pd107	&	$4.1 \times 10^{-10}$		&	$8.4 \times 10^{-10}$		&	$4.6 \times 10^{-10}$		&	$1.4 \times 10^{-9}$		&	$1.4 \times 10^{-9}$		&	N/A$^{d}$					\\  \hline
			\sm146	&	$3.9 \times 10^{-10}$		&	$6.3 \times 10^{-12}$		&	$3.4 \times 10^{-10}$		&	$8.5 \times 10^{-10}$		&	$1.2 \times 10^{-9}$		&	N/A$^{d}$					\\  \hline
			\hf182	&	$1.4 \times 10^{-10}$		&	$1.5 \times 10^{-9}$		&	$2.5 \times 10^{-10}$		&	$5.5 \times 10^{-10}$		&	$4.3 \times 10^{-10}$		&	N/A$^{d}$					\\  \hline
			\pu244$^{c}$ &	$2.0 \times 10^{-11}$	&	$2.2 \times 10^{-10}$		&	$3.7 \times 10^{-11}$		&	$8.1 \times 10^{-11}$		&	$6.3 \times 10^{-11}$		&	N/A$^{d}$					\\  \hline
			\hline
			Progenitor	&	6.5-$\msol$ SAGB$^{e}$	&	7.0-$\msol$ SAGB$^{e}$	&	7.5-$\msol$ SAGB$^{e}$	&	8.0-$\msol$ SAGB$^{e}$	&	8.5-$\msol$ SAGB$^{e}$	&	9.0-$\msol$ SAGB$^{e}$	\\  \hline
			\al26	&	$5.0 \times 10^{-6}$		&	$5.0 \times 10^{-6}$		&	$5.0 \times 10^{-6}$		&	$8.0 \times 10^{-6}$		&	$1.0 \times 10^{-5}$		&	$1.1 \times 10^{-5}$		\\  \hline
			\mn53	&	$0$							&	$0$							&	$0$							&	$0$							&	$0$							&	$0$							\\  \hline
			\fe60	&	$5.0 \times 10^{-6}$		&	$3.0 \times 10^{-6}$		&	$4.0 \times 10^{-6}$		&	$9.0 \times 10^{-6}$		&	$1.4 \times 10^{-5}$		&	$1.4 \times 10^{-5}$		\\  \hline
	\end{tabularx}
	{\raggedright
		{\it Note:  In addition to the cited CCSN yields from \citet{raus02}, \al26 and \fe60 yields from \citet{lc06} (11-120 $\msol$) were investigated as well.  These did not show any additional features beyond those shown with the \citet{raus02} yields. \\
		$^{a}$ - S15, S19, S20, S21, and S25 Models respectively, \citet{raus02} \\
		$^{b}$ - ``unchanged'' configuration, \citet{wana13} \\
		$^{c}$ - \pu244 yields calculated as outlined in \citet{fhe05} but using \citet{raus02} \hf182 yields \\
		$^{d}$ - $r$-process yields for ECSN are not available at present although \citet{wana13} stated that ECSN may produce some weak $r$-process elements \\
		$^{e}$ - \citet{doh13}} \\
	}
\end{table}
\end{center}

\subsection{Supernovae}
\label{sect:SN}

SNe include both CCSNe and TNSNe; CCSNe are the results of massive stars completing Fe/Ni fusion in their cores and collapsing under the influence of gravity whereas TNSNe result from runaway nuclear fusion in a C-O white dwarf near its Chandrasekhar limit.  Both types have similar explosive energies and modes of transporting ejecta.  However, although known to be sources of stable iron isotopes, TNSNe are calculated to produce relatively little \fe60, namely $\sim 2.3 \times 10^{-9} \ \msol$, according to the W7 Type Ia Model of \citet{nomo84}.  ECSNe form a special subcategory of CCSNe with significantly different radioactive isotope yields.  While the \fe60 yields for CCSNe and ECSNe are similar \citep[$\sim 10^{-6}-10^{-3} \ \msol$, ][]{raus02, lc06}, their yields for other isotopes (e.g., \al26 and \mn53) are vastly different \citep{wana13}.  For the purposes of this paper, ECSNe will refer to SNe from $8-10 \ \msol$ stars, CCSNe will refer to SNe with progenitor masses $> 10 \ \msol$, and SNe will refer to CCSNe, ECSNe, and TNSNe.  It should be noted that the yields for \pu244 were calculated using the same method from \citet{fhe05}; the proportions of $r$-process elements are generally consistent to that found in metal-poor globular clusters and in the Sun.  In this paper, however, the yields for \pu244 were based on the yields for \hf182 from \citet{raus02} using the ratios given in \citet{fhe05}.

SNe can show large variations in their isotope yields.  TNSN \fe60 yields show variations over several orders of magnitude ($\sim 10^{-18}-10^{-7} \ \msol$) due to variations in the number and location of ignition points and the transition from deflagration to detonation \citep{seit13}.\footnote{We adopt a fiducial TNSN \fe60 yield of $\sim 10^{-9} \msol$, which is consistent with the classic W7 result \citep{nomo84} and is larger than almost all \citet{seit13} models.}  CCSNe/ECSNe yields are highly dependent on a number of factors including when different layers are mixed.  This can be seen in the variations of yields from one mass to another \citep{raus02}.  The yields within a given mass are also subject to uncertainties in nuclear reaction rates (3-$\alpha$ and $^{12}$C($\alpha$,$\gamma$)$^{16}$O) which can lead to an almost order of magnitude shift in the production of \al26 and \fe60 \citep{tur2010}.  In \S \ref{sect:results}, Figure \ref{fig:dist}, we show the calculated distances with uncertainties indicated by dashed lines for a factor of 5 variation in the \fe60 yield for each CCSN/ECSN type.

Fiducial parameters for the explosions and the interstellar medium are chosen as follows:  We assume CCSNe and TNSNe deposit $E_{\rm CCSN \ \& \ TNSN} = 10^{51}$ ergs into their ejecta, while ECSNe deposit $E_{\rm ECSN} = 10^{50}$ ergs \citep{wana09}.  Because the Local Bubble shows evidence of multiple SN explosions, we will assume that if a SN were the source of the \fe60 signal, it would be the most recent SN, meaning the SN occurred in an already depleted ISM, but not as depleted as the current density of the Local Bubble (i.e., $n_{\rm Average \ ISM} = 1.0 \ {\rm cm}^{-3}> n_{\rm ISM} > n_{\rm Local \ Bubble} = 0.005 \ {\rm cm}^{-3}$).  Therefore, we estimated a SN would have occurred in an ISM of density, $n_{\rm ISM} = 0.1 \ {\rm cm}^{-3}$, temperature, $T = 8000$ K, and sound speed, $c_{s} = 10$ km s$^{-1}$ \citep[i.e., approximate values for the Local Cloud,][]{faj08}.

\subsection{Kilonovae}
\label{sect:KN}

KNe are thought to result from Neutron Star-Neutron Star (NS-NS) or in some cases Neutron Star-Black Hole (NS-BH) mergers \citep{lip1998,met2010,tan2013}.  For this paper, we only consider the KN explosion's lower-energy, spherical/torical ejection and not its highly beamed gamma-ray burst jet.  The rapid decompression and ejection of neutron-rich NS matter makes these events a natural site for the $r$-process \citep{ls74,ss82}.  While KNe are less energetic than SNe \citep[$E_{\rm KN} = 10^{49}$ ergs,][]{gor11}, we will consider a possible KN source of the \fe60 signal as occurring in the same ISM conditions as a SN.  However, given the axisymmetric nature of NS-NS mergers, we will not apply the same constraints to KNe as SNe, but will instead evaluate KNe with respect to isotope yields and frequency.

While there has been some modeling of KN yields, none we are aware of have specifically stated a yield for \fe60.  However, it is possible to determine an upper limit on the range for a KN.  In \citet{gor11}, they list mass fractions for every atomic number up to $\sim 200$ for a NS-NS merger with a total merger mass of 2.7 $\msol$.  If we assume all of the isotopes with $A = 60$ are in the form of \fe60 ($M_{\rm ej, total} = 10^{-3}-10^{-2}$, $X \textsubscript{\fe60} = 10^{-5}$) then the upper limit to the mass of ejecta in \fe60 is 10$^{-7}$ $\msol$.

\subsection{Super-Asymptotic Giant Branch Stars}
\label{sect:SAGB}

SAGBs ($6.5-9 \ \msol$) are post-main sequence stars that produce large amounts of dust \citep[see, e.g.,][]{vent12} and have strong winds ($\sim 30$ km s$^{-1}$) capable carrying dust great distances.  SAGBs produce 10$^{-6}-10^{-5}$ $\msol$ of \fe60 \citep{doh13}, but are distinguishable from SNe in that they produce practically no \mn53 \citep{wass06,fimi14}.  We note that SAGB yields are subject to an uncertainty in the onset of the super-wind phase \citep{doh13}; a delayed onset results in generally increased yields.  The implication for distance is shown with dashed error bars on SAGB results in \S \ref{sect:results}, Figure \ref{fig:dist}.  In contrast to SNe, we do not expect a SAGB wind to affect the density of the Local Bubble appreciably, and we assume that a SAGB source for the \fe60 signal would have occurred in an ISM like that found in the Local Bubble today (i.e., $n_{\rm ISM} = 0.005 \ {\rm cm}^{-3}$, temperature, $T = 10^{6}$ K, and sound speed, $c_{s} = 100$ km s$^{-1}$).  Finally, we assume the initial velocity of the SAGB grains to be:  $v_{\rm grain, 0} = 30 \ {\rm km \ s^{-1}}$.

\subsection{Dust Transport to the Solar System}
\label{sect:dustparam}

Regardless of the source, any \fe60 arriving in the Solar System will need to be in the form of dust.  \citet{faj08} showed that the solar wind will keep any gaseous isotopes from reaching the Earth (unless a SN is sufficiently close, but this will be used as a constraint later in \S \ref{sect:SNprof} and \S \ref{sect:results}).  We assume that the dust grains are spherical and select as our fiducial values for dust grains:  density, $\rho_{\rm grain} = 3.5 \ {\rm g \ cm^{-3}}$ (an average value for silicates), radius, $a = 0.2 \ \mu{\rm m}$ (this selection is based on discussion in \S \ref{sect:dust}), and voltage, ${\cal V} = 0.5 \ {\rm V}$.  Departures from these values will be specifically stated.

\section{Formalism}
\label{sect:formalism}

In order to identify the most likely progenitor, we will attempt to constrain the source and its allowable distances using the measured \fe60 fluence and calculated yields.  In the future, with additional measurements of other isotopes, we can use the other isotope yields to constrain the source using the observed isotope to \fe60 ratio.

Several previous works have presented the formalism for calculating deposited material from a SN \citep{efs96,fe99,fhe05}. 
These works focused primarily on short ranges (SN distances, $D \sim 10$ pc)
and on the isotope \fe60.
For such short distances, the  losses due to decay of live radionuclides en route from the SN to Earth amount to $\lesssim 1\%$ and can be ignored.
At greater ranges ($D \sim 100$ pc) and for shorter-lived isotopes 
(in particular, \al26 with $\tau_{1/2,}\textsubscript{\al26}$ = 0.717 Myr) decays en route become a significant issue.  
Accounting for this, the observed fluence today, ${\cal F}_{{\rm obs}, i}$, for each isotope, $i$, in atoms per area on the surface of the Earth within a given substance (e.g., crust, sediment, etc.) becomes:
\begin{equation}
\label{eq:Ntotobs}
	{\cal F}_{{\rm obs}, i} = \left( \frac{1}{4} \right) \left( \frac{M_{{\rm ej}, i}}{4 \pi D^{2} A_{i} m_{u}} \right) U_{i} \ f_{i} \ e^{- \left( \tarr + \tsn \right) /\tau_{i}} 
\end{equation}
where $M_{{\rm ej},i}$ is the mass of the ejecta by isotope, $D$ is the distance from the progenitor to Earth, $A_{i}$ is the atomic number of the isotope, $m_{u}$ is the atomic mass unit, $\tarr$ is the time from today since the ejecta arrived at Earth, and $\tsn$ is the time the ejecta traveled from the source to Earth.  Also, $\tau_{i}$ is the mean lifetime of the isotope ($\tau_{i} = \tau_{1/2, i}/\ln 2 $).

The uptake, $U_{i} \equiv ({\rm Amount \ Collected})/({\rm Amount \ Deposited})$, is the fraction of the isotope deposited on a surface that is collected by that material.  The quantity is dimensionless, and ranges from 1 (the material collects 100\% of deposited element) to 0 (the material collects 0\% of deposited element).\footnote{It is possible to have $U_i > 1$ if, for example, a marine sample can chemically scavenge the element of interest so efficiently that it collects more than the amount deposited in the water column directly over the sample.}  It is further discussed in \S \ref{sect:uptake}.  Additionally, the dust fraction, $f_{i}$, is the amount of the isotope in the form of dust that arrives at Earth (\S \ref{sect:dust}).  It is similar to uptake in that it is also dimensionless and ranges from 1 (all of the isotope is in the form of dust and reaches Earth) to 0 (none of the isotope is dust and/or reaches Earth).  There is a factor of $(1/4)$ from the ratio of the Earth's cross sectional to surface areas, because it is assumed material is distributed evenly over Earth's entire surface through collisional accretion only.  Equation (\ref{eq:Ntotobs}) also assumes an isotropic dispersal of material from the source ($4 \pi D^{2}$ factor for spherical distribution), that no additional isotopes are created after the ejection from the progenitor, and that the ejected material passes through a homogeneous ISM.

There are three other fluence quantities that appear in the literature:  ``decay-corrected'' fluence, ``surface'' fluence, and ``interstellar'' fluence.  The decay-corrected or arrival fluence, ${\cal F}_{{\rm arr}, i}$, is the total number of atoms per area that would have been measured at the time the signal arrived.  It is calculated by correcting our previous description of fluence (see Equation (\ref{eq:Ntotobs})) for radioactive decay since the isotope was deposited ${\cal F}_{{\rm arr}, i} = e^{\tarr /\tau_{i}} {\cal F}_{{\rm obs}, i}$.  The surface fluence or global mean fluence, ${\cal F}_{{\rm surface}, i}$, is the total number of atoms per area that arrive at the surface of the Earth regardless of what substance they might be incorporated into, and is found by dividing the decay-corrected fluence by the uptake (i.e., ${\cal F}_{{\rm surface}, i} = {\cal F}_{{\rm arr}, i}/U_i$).  It will be used in \S \ref{sect:data} and \S \ref{sect:results} and will be specifically stated when used.  The interstellar fluence, ${\cal F}_{{\rm interstellar}, i}$, also appears in the literature \citep[e.g.,][]{fit08,cook09}, namely the number of atoms per area on the surface of the spherical shock front.  It is related to the surface fluence by a factor of 4, the ratio of Earth's cross section to surface area (i.e., ${\cal F}_{{\rm interstellar}, i} = 4{\cal F}_{{\rm surface}, i}$).  Interstellar fluence will not be used in this paper, but the reader should be aware of the distinction when reviewing the literature.

In order to find the time delay, $\tsn$, from ejection to deposition on Earth, we must account for the propagation of the ejection through the intervening ISM, which depends on the progenitor.  SNe transmit material via an explosive shock, whereas SAGBs use a wind-driven ejection.  In the case of SNe, we will assume that the shock has transitioned from the free-expansion phase into the adiabatic/energy-conserving phase.  For SAGBs, we will assume dust has been blown by winds from the star and experiences drag as it travels to Earth.

\subsection{SN Expansion Profile and Constraints}
\label{sect:SNprof}

For SNe in the adiabatic/energy-conserving phase, the shock follows a self-similar or Sedov-Taylor expansion profile.  With the explosion at time $t=0$, a shock is launched with radius, $\Rs$, at elapsed time, $t$, given by:
\beq
\label{eq:sedov}
	\Rs = \xi_0 \pfrac{\Esn t^2}{\rho_{\rm ISM}}^{1/5} 
\eeq
for a SN explosion depositing energy $\Esn$ into the ejecta,
propagating into a local interstellar medium of density $\rho_{\rm ISM} = m_u n_{\rm ISM}$.
The quantity $\xi_{0}$ is a dimensionless constant that is of order unity for $\gamma = 5/3$ using the derivation in \citet{zr}.  
Thus the time interval to traverse distance, $D$, is:
\begin{equation}
\label{eq:tsn}
	t_{\rm travel, SN} = \left( \frac{D}{\xi_{0}} \right)^{5/2} \left( \frac{\rho_{\rm ISM}}{\Esn} \right)^{1/2}
\end{equation}
where we have assumed a uniform ISM density.  While we know this is a crude approximation \citep[see e.g.,][]{abt2011}, deviations from the uniform case would be encoded into the signal and could be determined if the signal is time-resolved \citep[for examples of non-uniform media, see][]{bk1994}.  We use the uniform ISM case as a baseline.

The density versus radius profile for a SN signal may be approximated by a ``saw-tooth'' profile.  As will be described in greater detail in \S \ref{sect:raddist}, \S \ref{sect:timesig}, and Appendix \ref{sect:appendix}, measurements in sediment open the possibility to making time-resolved fluence measurements.  A saw-tooth pattern gives a better approximation of the more exact Sedov solution than a uniform shell profile.  The saw-tooth pattern reaches its maximum density value at the outer edge of the shock, then decreases linearly to a fraction of the total shock radius, $\epsilon$, where the density is zero from that point to the center of the remnant.  See Appendix \ref{sect:appendix} for a comparison of the exact Sedov, saw-tooth, and uniform shell profiles.

Possible SNe will be constrained by an inner ``kill'' distance and an outer ``fadeaway'' distance.  The kill distance is the range at which a SN can occur and create extinction-level disruptions to the Earth's biosphere.  The primary mechanism for a SN to accomplish this is for ionizing radiation (i.e., gamma-rays, hard X-rays, and cosmic-rays) to destroy O$_{3}$ and N$_{2}$ in the atmosphere producing nitric oxides (NO$_{y}$) and leaving the biosphere vulnerable to UVB rays from the Sun \citep[first described by][described in detail by \citealt{rud74}, and updated by \citealt{es95}]{ss66}.  This can be accomplished either by direct exposure (i.e., an X-ray flash from the SN) or by a ``descreening'' boost in cosmic-ray flux.\footnote{Note, our kill distance does not include the case of a gamma-ray burst given their narrowly-beamed emission.}  \citet{gehr03} calculated a kill distance $R_{\rm kill} \lesssim 8$ pc for the direct exposure case using the galactic gamma-ray background and SN rates, although, as pointed out by \citet{melt11}, this is probably an underestimation based on more recent rate estimations.  This work was expanded upon by \citet{ezj07} and \citet{thom08} to include X-rays and showed that for exposure durations up to $10^{8}$ s, the effects on the biosphere were the same and that the critical value for an extinction-level SN event was an energy fluence of $10^{8} \ {\rm ergs \ cm^{-2}}$ (not to be confused with the description of fluence used throughout the rest of this paper).  As noted in \citet{melt11}, these direct exposure calculations use SN rate and photon and cosmic-ray emission information that are improving, but still subject to large uncertainties.

Here we calculate the kill distance using the descreening case described in \citet{faj08}; it yields the same range as the direct exposure calculations, and is scalable to the energy of the SN, $\Esn$.  The descreening kill distance is the range at which a SN can occur and its shock will penetrate the Solar System to within 1 AU of the Sun.  It is determined by setting the solar wind pressure, $P_{\rm SW}$, equal to the pressure of the SN shock \citep[see, e.g.,][]{faj08}.  In this case, the pressure from the SN has little effect on the Earth, but by pushing back the solar wind, the Earth is inside the SN remnant and now directly exposed to the SN cosmic-rays that would normally diffuse out over $10^{4}$ yr \citep{fuj10} in addition to an increased galactic cosmic-ray background.  In-turn, these destroy O$_{3}$ and N$_{2}$ in the atmosphere, just as in the direct exposure case, in addition to increased radionuclide deposition \citep{melt11}.  Using our fiducial SN values, we find:
\beq
	R_{\rm kill} = 10 \ {\rm pc} \ \left( \frac{\Esn}{10^{51} \ {\rm ergs}} \right)^{1/3} \left( \frac{2 \times 10^{-8} \ {\rm dyne \ cm^{-2}}}{P_{\rm SW}} \right)^{1/3}
\eeq

The fadeaway distance is the range at which the SN shock dissipates and slows to the sound speed of the ISM.  Because of uncertainty in when SN dust decouples from the rest of the shock, the fadeaway distance is not an absolute limitation like the kill distance, but can serve as a guide to the likelihood of a progenitor.  Using the derivation from \citet[][Eq. 39.31]{drain11}, we find:
\beq
	R_{\rm fade} = 160 \ {\rm pc} \ \left( \frac{\Esn}{10^{51} \ {\rm ergs}} \right)^{0.32} \left( \frac{0.1 \ {\rm cm^{-3}}}{n_{\rm ISM}} \right)^{0.37} \left( \frac{10 \ {\rm km \ s^{-1}}}{c_{s}} \right)^{2/5}
\eeq

\subsection{SAGB Expansion Profile and Constraints}
\label{sect:SAGBprof}

In the case of SAGBs, we assume the dust is ejected radially and that the distance traveled by SAGB dust is determined only by a drag force, $F_{\rm drag}$ (magnetic forces will be considered later in this section as a constraint).  Using the description in \citet{drain79}, the drag force due to only collisional forces (in cgs units) is:\footnote{The Coulomb force can be large, however, it is not for our selected grain parameters, so we neglect Coulomb forces \citep[see constraints for $(\phi ^{2} \ln \Lambda)$ in][\S 26.1.1]{drain11}.}
\beq
\label{eq:draine}
	F_{\rm drag} = 2 \pi a^{2} kT \left( \sum_{j} n_{j} \left[ G_{0}(s_{j}) \right] \right)
\eeq
with
\beq
	G_{0}(s) \approx \frac{8s}{3 \sqrt{\pi}} \left( 1+\frac{9 \pi}{64} s^{2} \right)^{1/2}, \; \; s_{j} \equiv \left( m_{j} v^{2}/2kT \right)^{1/2} \,
\eeq
where $j$ is the respective species in the ISM (we consider only ionized H; He and free electrons will be neglected), $m_{j}$ is the particle mass of that respective species, $k$ is the Boltzmann constant, $T$ is the temperature of the ISM, and $v$ is the velocity of the particle relative to the medium.  For small $v$ (i.e., $v_{\rm grain} \lesssim 100$ km s$^{-1}$) the first term in $G_{0}$ will dominate, leaving $G_{0}(s) \approx 8s/(3 \sqrt{\pi})$.  Making these simplifications, we find:
\beq
	F_{\rm drag} = 2 \pi a^{2} kT n_{\rm ISM} v_{\rm grain} \left( \frac{8}{3 \sqrt{\pi}} \right) \left( \frac{m_{p}}{2kT} \right)  ^{1/2} = m_{\rm grain} \ddot{R}_{\rm SAGB}
\eeq
where $\Ra$ is the distance the dust grain has traveled from the SAGB, and $v_{\rm grain} \equiv \dot{R}_{\rm SAGB}$.  Integrating twice and setting $\Ra$ equal to the traverse distance from the progenitor to Earth, $D$, gives the transit interval:
\beq
	t_{\rm travel, SAGB} = -\zeta_{0} \left( \frac{a \rho_{\rm grain}}{c_{s} \rho_{\rm ISM}} \right) \ln \left( 1 - \frac{1}{\zeta_{0}} \frac{c_{s} \rho_{\rm ISM}}{a \rho_{\rm grain}} \frac{D}{v_{\rm grain, 0}} \right)
\eeq
where $\zeta_{0}$ is a dimensionless constant and of order unity for $\gamma = 5/3$, $c_{s}$ is the sound speed in the ISM, and $v_{\rm grain, 0}$ is the initial velocity of the dust grain when it leaves the SAGB.

We approximate the density profile for a SAGB signal by a uniform shell or ``top-hat'' shape (we will use ``top-hat'' profile to avoid confusion with the SN profile discussion).  This corresponds to a uniform, steady wind.  The top-hat pattern reaches its maximum density value at the leading edge of the signal, retains this value for the duration of the signal, and afterwards the density returns to zero.  The SAGB phase is characterized by thermal-pulsing of the star's envelope.  Since the duration of each pulse ($\Delta t_{\rm pulse} \sim 1$ yr) and interval between pulses ($\Delta t_{\rm inter} \sim 100$ yr) are much shorter than the SAGB phase ($\Delta t_{\rm SAGB} \sim 100 \ {\rm kyr} \gg \Delta t_{\rm inter} > \Delta t_{\rm pulse}$) we assume that the amount of ejected material is approximately constant for the duration of the SAGB phase \citep[see][]{siess10}.  Furthermore, we assume all parts of the signal experience the same forces from the SAGB to the Earth, so that the duration of the signal remains the same (i.e., $\Delta t_{\rm signal, SAGB} = \Delta t_{\rm SAGB} = 100$ kyr).

Because SAGB winds would not be as devastating to the Earth as a SN shock, we forego establishing an inner kill distance, but establish two outer distances:  the drag stopping distance, $R_{\rm drag}$, and the magnetic deflection distance, $R_{\rm mag}$.  The distance, $R_{\rm drag}$, is the range of the SAGB dust grains' $e$-folding velocity, and the $R_{\rm mag}$ is the range at which deflection of the dust grain's trajectory by the ISM's magnetic field becomes significant.  Using the derivations from \citet{murr04}, we find:
\beq
	R_{\rm drag} = 93 \ {\rm kpc} \left( \frac{\rho_{\rm grain}}{3.5 \ {\rm g \ cm^{-3}}} \right) \left( \frac{a}{0.2 \ \mu{\rm m}} \right) \left( \frac{0.005 \ {\rm cm^{-3}}}{n_{\rm ISM}} \right) \left( \frac{10^{6} \ {\rm K}}{T} \right)^{1/2} \left( \frac{v_{\rm grain, 0}}{30 \ {\rm km \ s^{-1}}} \right)
\eeq
\beq
	R_{\rm mag} = 0.02 \ {\rm pc} \left( \frac{\rho_{\rm grain}}{3.5 \ {\rm g \ cm^{-3}}} \right) \left( \frac{0.5 \ {\rm V}}{\cal V} \right) \left( \frac{5 \ \mu{\rm G}}{B} \right) \left( \frac{v_{\rm grain, 0}}{30 \ {\rm km \ s^{-1}}} \right) \left( \frac{a}{0.2 \ \mu{\rm m}} \right)^{2}
\eeq
The implications of these limits will be discussed in greater detail in \S \ref{sect:results}.

\subsection{The Radioactivity-based Distance to the Explosion}
\label{sect:raddist}

To estimate the distance, $D$, to the explosion, 
we wish to invert
Equation (\ref{eq:Ntotobs}). 
When the transit time (the time for the shock to travel from the source to Earth) is negligible ($\tsn \ll \tau_i$), the procedure
is straightforward, since the only distance dependence is the
inverse square dilution of the ejecta, and so
$D \propto 1/\sqrt{{\cal F}_{{\rm obs}, i}}$.\footnote{This ``radioactivity distance'' is analogous to the usual luminosity distance:  the yield plays the role of luminosity, and radioisotope fluence the role of flux \citep{ltf06}.}
This has been assumed in work to date.
However, if $D$ is sufficiently large,
then via Equation (\ref{eq:sedov}), the distance-dependent transit time 
can become important and must be included in solving
Equation (\ref{eq:Ntotobs}); we have done this in all of our results.

Another effect occurs when the radioisotope signal is sampled
sufficiently finely to resolve the time history of the deposition signal.
This occurs when the signal width (the time from the arrival of the signal's leading edge to the departure of the signal's trailing edge)
is larger than the sampling time resolution ($\dtsignal > \tres$).  
In this case, the total radioisotope signal, summed over all time
bins, should be used in solving the distance via Equation (\ref{eq:Ntotobs});
and as we show in \S \ref{sect:timeresolv} below, 
the width of $\dtsignal$ for a SN probes independently the explosion distance.
However, the available \citet{knie04} data has a time sampling
of $\tres \sim 880$ kyr, and shows no evidence for a signal
that is extended in time.  Thus we infer that the signal width
$\dtsignal \la \tres$, and indeed we find
$\dtsignal \la 880$ kyr for most of our possible progenitors.
In addition, when solving for distance using the Knie results, we assumed the signal arrived halfway through the sample.  Therefore, half of the sampling width is used as a median value rather than assuming the signal arrives right as the sampling window begins or just before it ends.

\subsection{Expected Behavior of Time-Resolved Signals}
\label{sect:timesig}

Although the time resolution of the deep-ocean \fe60 crust measurements in \citet{knie04} data preclude resolution of the time structure of the radioisotope deposition, measurements
in sediments can achieve much better time resolution, and so it is of interest to explore the
time dependence of the explosion signal.
Such work was pioneered by \citet{ammo91}
in the context of the \be10 anomalies $\sim 35$ and 60 kyr ago in 
Antarctic ice cores \citep{rais87}.

In developing a model for deposition, we examine a Sedov-Taylor profile for a SN shock (a similar examination could be done for the SAGB case if we had more detailed description of its signal width dependence), which implies an energy-conserving (adiabatic) evolution.
This also means the shock will remain self-similar as it progresses and that the majority of the material is concentrated near the leading edge.  
Although the remnant density profile changes once the remnant transitions to the radiative/momentum-conserving phase, we have chosen to maintain the Sedov profile.  We did this, firstly, because the profiles are similar in shape \citep[the radiative profile is a thicker shell profile, see, e.g.,][Fig. 17.4]{shu92}.  Secondly, because the dust will decouple from the gas at some point, either when the shock meets back pressure from encountering the solar wind or at the transition from the adiabatic to the radiative phase when the shock loses its internal radiation pressure \citep{drain11}.  The exact nature of this decoupling and resulting profile would require detailed calculations that are beyond the scope of this paper.

We assume that the explosion ejecta are well-mixed within the 
swept-up matter, so that the ejecta density profile
follows that of the blast itself.
As described in Appendix \ref{sect:appendix}, we approximate
the Sedov density profile as a ``saw-tooth'' that drops
linearly from a maximum behind the shock radius $\Rs$ to 
zero at an inner radius $r_{\rm in} = (1-\epsilon) \Rs$,
with $\epsilon \approx 1/6$.
We note that the leading edge of the blast  
from an event at distance $D$ arrives
at a time $\tsn$ since
the explosion which corresponds to
geological time $\tarr$; this is given by $D = \Rs(\tinit)$,
whereas the trailing edge of the shell arrives at 
time $\tfin$ given by $D = (1-\epsilon) \Rs(\tfin)$.
Thus we have
$\tfin=\tinit/(1-\epsilon)^{5/2}$.

With these assumptions we can model
the global-averaged flux time profile, $\mathbb{F}$.
For radioisotope $i$,  we have:
\beq
\label{eq:timeprof}
\mathbb{F}_i(D,t) =  \pfrac{t}{\tinit}^{-11/5} \ 
 \left[ 
   \frac{(t/\tinit)^{-2/5}-(\tfin/\tinit)^{-2/5}}{1-(\tfin/\tinit)^{-2/5}}
 \right] 
 \ \mathbb{F}_i(D,\tinit)  
\eeq
as shown in Appendix \ref{sect:appendix};
note that here all times are geological and thus increase towards the past.\footnote{In Equation (\ref{eq:sedov}) time increases towards the future, but note that elsewhere, unless explicitly stated, times are used in the geological sense, and thus increase towards the past.}  This describes a cusp-shaped decline from an initial flux:
\beqar
\mathbb{F}_i(D,\tinit) &= & \frac{2{\cal F}_{1}}{5\tinit}
 = \frac{3}{40\pi} \frac{\gamma+1}{\gamma-1} \frac{M_{{\rm ej}, i}/m_i}{D^2} \
  \pfrac{\xi_0^5 E_{\rm SN}}{m_p n_{\rm ISM} D^5}^{1/2} \\
 & = & 9.5 \times 10^5 \ {\rm atoms \ cm^{-2} \ kyr^{-1} }  \\
\nonumber
  & & \ \times \ \pfrac{0.1 \ \rm cm^{-3}}{n_{\rm ISM}}^{1/2}
   \ \pfrac{E_{\rm SN}}{10^{51} \ \rm ergs}^{1/2} 
   \ \pfrac{M_{{\rm ej}, i}}{3 \times 10^{-5} \ \msol}
   \ \pfrac{60}{A_{i}}
   \ \pfrac{100 \ \rm pc}{D}^{9/2}
\eeqar
where the numerical values are $m_i = 60m_u$
and the yield is appropriate for \fe60.

To test the profile in Equation (\ref{eq:timeprof}), one can fit observed
time-resolved data to this form, letting $\tinit$ and $\tfin$,
and $\mathbb{F}_i(D,\tinit)$ be free parameters.
For the Sedov profile, the time endpoints should
obey $\tfin=\tinit/(1-\epsilon)^{5/2}$, which provides
a consistency check for the Sedov (adiabatic) approximation.
Moreover, as we show in \S \ref{sect:timeresolv},
the interval $\tfin-\tinit$ provides an independent
measure of the explosion distance.

If the radioisotope abundance is sampled over
a time interval $[t_1,t_2]$, then 
the fluence (without radioactive decays) 
will be the integral of the surface flux 
over this interval:
${\cal F}_i(t_1,t_2) = \int_{t_1}^{t_2} \mathbb{F}_i(t) \ dt$, 
where we have suppressed the dependence on time-independent
parameters such as distance.
If the observed time resolution $\tres$ is small compared to
$\tinit$ and $\tfin$, 
then the fluence profile
${\cal F}_i(t-\tres/2,t+\tres/2) \approx \mathbb{F}_i(t) \ \tres$
will recover the flux history.

So far we have calculated the observed fluence without the effect of decay.
Since all the atoms were created at the same time, 
the observed fluence is reduced by a factor
$e^{-(\tarr+\tsn)/\tau_i}$.
The effects of uptake (\S \ref{sect:uptake}) 
and dust depletion (\S \ref{sect:dust})
introduce further factors of
$U_i$ and $f_i$.  We thus arrive at the observed fluence:
\beq
\label{eq:Nresolved}
{\cal F}_{{\rm obs}, i}(t_1,t_2) = U_i f_i e^{-(\tarr+\tsn)/\tau_i}
  \int_{t_1}^{t_2} \mathbb{F}_i(t) \ dt \ . 
\eeq
One can show that the total integrated fluence
${\cal F}_{{\rm obs}, i}(\tinit,\tfin)$ takes precisely the
value in Equation (\ref{eq:Ntotobs}). 
This reflects the number conservation (aside from decays)
of the atoms in the SN  ejecta.
This also implies that, in a time-resolved measurement,
the total fluence is conserved,
which means that the area under a fluence vs. time curve
will be constant for fixed explosion parameters.
This implies that fluence measurements of \fe60/Fe will show lower values when measured over fewer bins, and finer sampling will show higher fluence over more bins (see Figure \ref{fig:timeres}).

\subsection{A Resolved Signal Timescale Probes the Distance to the Explosion}
\label{sect:timeresolv}

A time-resolved signal not only encodes information about
the shape of the blast density profile, but also
about the distance.  The relation $D = (1-\epsilon) \Rs(\tfin)$
also allows us to write $\tfin=\tinit/(1-\epsilon)^{5/2}$ and thus that:
\beq
\dtsignal = \tfin - \tinit \equiv \alpha \tinit
\eeq  
where we define a dimensionless parameter, $\alpha$, that relates the signal width in terms of the arrival time.  For our profile, $\alpha = (1-\epsilon)^{-5/2}-1 \approx 0.577$.
We see that the radioisotope width grows in proportion to the
blast transit time to Earth, $\tsn$, which itself depends on distance.
Thus a measurement of $\dtsignal$ from a time resolved 
radioisotope signal gives an measure of
the explosion distance.  Within this Sedov model, 
$D = \Rs(\tinit)$, and so we can solve for
the distance based on ``blast timing:'' 
\beqar
\label{eq:Dblast}
D & = & \xi_0 \pfrac{\Esn \dtsignal^2}{\alpha^2 \rho}^{1/5} \\
  & = & 65 \ {\rm pc} 
    \ \pfrac{\dtsignal}{100 \ \rm kyr}^{2/5}
    \ \pfrac{\Esn}{10^{51} \ \rm ergs}^{1/5}
    \ \pfrac{0.1 \ \rm cm^{-3}}{n_{\rm ISM}}^{1/5}
\eeqar

This distance measure is independent of the ``radioactivity distance''
and its associated uncertainties, notably due to
uptake, dust fraction, and radioisotope yields.
Moreover, as characteristic for Sedov blast waves,
the blast-timing distance in Equation (\ref{eq:Dblast}) scales as small powers of 
the timescale, as well as the energy and density.
This will weaken the uncertainties in distance estimate.

\begin{figure}
	\begin{center}
		\epsfig{file=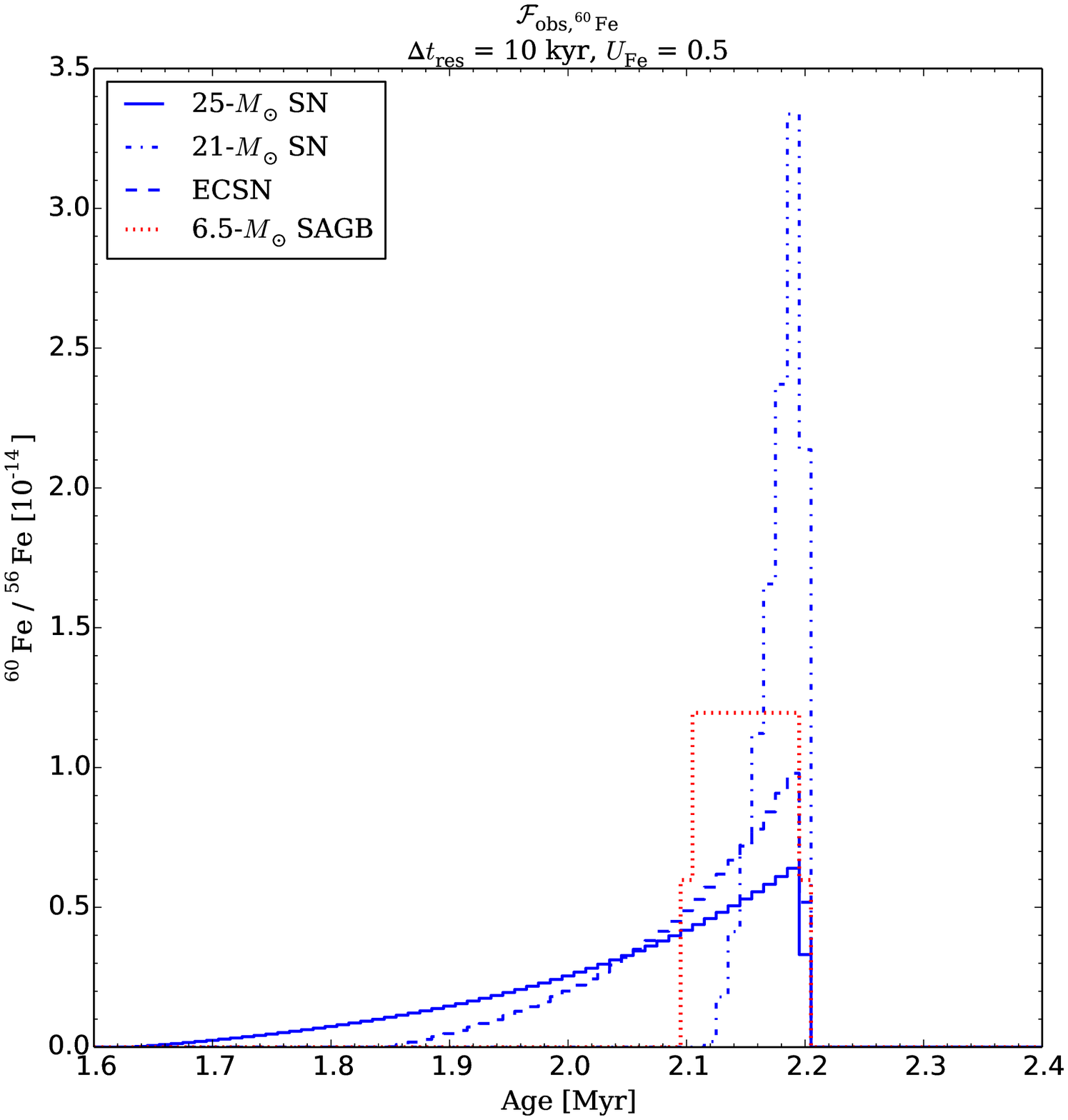,width=\textwidth}
		\caption{\it Sample time-resolved calculations of the observed fluence, ${\cal F}_{\rm obs, }\textsubscript{\fe60}$, for three SNe and a SAGB.  Each progenitor is at a different distance:  25-$\msol$ SN at 130 pc, 21-$\msol$ SN at 59 pc, ECSN at 67 pc, and 6.5-$\msol$ SAGB at 79 pc.  Of note, each of these progenitors would produce the same measured \fe60 signal by \citet{knie04,fit08}, but with a finer time-resolution (10 kyr in this case), the shape of the signal is readily discernible.  Also note that since the plots produce the same observed fluence, the areas under the curves are the same.
		\label{fig:timeres}
		}
	\end{center}
\end{figure}

Having two independent distance estimates 
allows a consistency check for the model.
Alternatively, if we adopt one of the distance estimates
as the correct value,
we can deduce the parameters in the other.
For example, adopting the blast-timing distance
we can use the observed fluence and solve for the product of radioisotope yield, uptake, and dust fraction:
$M_{{\rm ej}, i} U_i  f_i \propto D^2 \ {\cal F}_{{\rm obs}, i}$.
Given a geophysical estimate of uptake, this allows for a measure of the yield and thus a direct probe
of the nucleosynthesis output and thus the nature of the explosion (see Figure \ref{fig:timeres} for examples).

\section{Deposit Factors}

The delivery of astrophysical debris to the Solar System and its incorporation into geological and lunar samples is clearly complex.  In this section we consider several factors we are aware of that can have a substantial influence on the observed signals.

\subsection{Uptake}
\label{sect:uptake}

Uptake in the Fe-Mn crust involves a complex chemical process that incorporates material into the crust with a low accumulation rate $\sim 2$ mm Myr$^{-1}$.  Usual deep-ocean sediments, on the other hand, do not make such a geochemical selection, and have greater accumulation rates $\sim 3-4$ mm kyr$^{-1}$ \citep{feige12}.  The Fe uptake factor was calculated by \citet{knie04} using the relative concentrations of Fe and Mn in water and the Fe-Mn crust and the uptake of Mn (4\%), leading to an estimate for the Fe uptake, $U_{\rm Fe} = 0.6\%$.  However, recent studies have suggested that $U_{\rm Fe} = 0.5 - 1$ \citep{be11, feige12}.  Using the smaller estimate of $U_{\rm Fe}$, \citet{knie04} calculated a SN distance of $D \approx 40 \ {\rm pc}$; a reasonable distance considering the Local Bubble is $\sim 200$ pc in diameter in the Galactic plane extending 600 pc perpendicular to the plane \citep{fuchs06} and superbubbles in the Large Magellanic Cloud (LMC) are typically $\sim 100-200$ pc in diameter \citep[for a single round of star formation,][]{chu08}.  Changing the uptake factor has the immediate effect of changing the implied distance to the explosion.  This can be roughly understood if we ignore the effect of the debris decays in transit, in which case the signal follows the inverse square law and we have $D \propto \sqrt{U_{i}/{\cal F}_{{\rm obs}, i}}$.  The effect of decays en route softens this dependence somewhat.

If $U_{\rm Fe}$ is an order of magnitude larger, the implied distance increases by a factor $\ga 2$.  
With a Fe-Mn crust uptake factor of $U_{\rm Fe} = 0.5-1$, the implied distances are around $D \sim 200$ pc.  This seems an unlikely distance, given that the Solar System is roughly in the center of the Local Bubble \citep[see][Fig. 2]{bb02}, and a SN would have had to occur outside the Local Bubble in order to produce the signal (assuming the progenitor is in the Galactic plane).

However, implicit in the \citet{knie04} calculation is the assumption that the dust fraction, $f_{\rm Fe} = 1$.  As we will show in \S \ref{sect:dust}, this is most likely not the case, and the combination of a higher uptake value (we chose $U_{\rm Fe} = U_{\rm Al} = 0.5$) with a smaller dust fraction ($f_{i} \ll 1$) can still yield reasonable progenitor distances.

\subsection{Dust Condensation}
\label{sect:dust}

It was shown in \citet{faj08}, that for a SN further than $D \sim 10$ pc, the solar wind would keep the SN blast plasma outside of 1 AU, and thus the Earth will not find itself inside gas-phase SN debris.  
However, refractory SN ejecta will be condensed into dust grains.  As discussed in \citet{af11}, we expect these grains to be entrained in the SN blast as it reaches the heliosphere, but then decouple at the SN-solar wind shock and move essentially ballistically through the inner solar system.  Once the dust decouples from the gas in the shock, it can travel great distances.  At this point, both SNe and SAGBs behave the same, as they are subject to the same drag stopping distance, $R_{\rm drag}$ discussed in \S \ref{sect:SAGBprof}.
This is more than sufficient to reach the Earth in spite of the solar wind, and indeed should carry dust grains beyond the SN remnant when it finally comes to rest.  Thus, for the $D > 10$ pc events of interest, the amount of any radioisotope $i$ that comes to Earth will be proportional to the fraction, $f_i$, of the isotope that reaches Earth via dust, as seen in Equations (\ref{eq:Ntotobs}) and (\ref{eq:Nresolved}).

Determining $f_{i}$ requires examining a number of factors (the results are summarized in Table \ref{tab:dust}): \\
1)  How much of the isotope condenses into dust at departure from source? \\
2)  How much of that dust survives the interstellar journey from the source to the Solar System? \\
3)  How much of the remaining dust can filter through the heliopause and enter the Solar System? \\
4)  How much of the filtered dust can overcome the solar wind/radiation pressure and reach the Earth's orbit? \\
\begin{center}
\begin{sidewaystable}
\caption{Summary of Relevant Dust Filtering Processes
\label{tab:dust}
}
	\begin{tabularx}{\textwidth}{ |c|c|c|C|C|C|C|C| } \hline
		Resulting Dust Fraction & Repository Material & Density & Elemental fraction in material & Fraction condensed into dust & Fraction surviving interstellar passage & Fraction passing into heliosphere & Fraction reaching Earth's orbit \\ \hline \hline
		\multirow{2}{*}{$f_{\rm Fe, SN} \approx 0.01$}
			&	Fe						& 7.9 & 0.1$^{a}$  & 1$^{c}$ & 1$^{a}$    & 0.1$^{a,d}$  & 1$^{g}$ \\
			&	FeS						& 4.8 & 0.9$^{a}$  & 1$^{c}$ & 0.5$^{a}$  & 0$^{a,d}$    & 1$^{g}$ \\ \hline 
		\multirow{3}{*}{$f_{\rm Fe, SAGB} \approx 0.2$}
			&	Fe						& 7.9 & 0.48$^{b}$ & 1$^{c}$ & 1          & 0.1$^{a,d}$  & 1$^{g}$ \\
			&	[Fe, Mg]$_{2}$SiO$_{4}$	& 3.5 & 0.43$^{b}$ & 1$^{c}$ & 1          & 0.25$^{d,e}$ & 1$^{g}$ \\
			&	[Fe, Mg]SiO$_{3}$		& 3.5 & 0.09$^{b}$ & 1$^{c}$ & 1          & 0.25$^{d,e}$ & 1$^{g}$ \\ \hline
		{$f_{\rm Al, SN} \approx 0$}
			&	Al$_{2}$O$_{3}$			& 4	  & 1$^{a}$    & 1$^{c}$ & 0.01$^{a}$ & 0$^{a,d}$    & 1$^{g}$ \\ \hline
		{$f_{\rm Al, SAGB} \approx 1$}
			&	Al$_{2}$O$_{3}$			& 4	  & 1$^{a}$    & 1$^{c}$ & 1          & 1$^{d,f}$    & 1$^{g}$ \\ \hline
	\end{tabularx}
	{\raggedright
		{\it Density is given in g cm$^{-3}$. \\
		Mass fractions from each filter process are determined using the relative size number distributions, $N(a)$, given in \citet{silv10} and \citet{mrn77}, multiplying by volume, $V(a)$, and density, $\rho$, and integrating over the relevant radii $[a_{max}$, $a_{min}]$.  This method is described in \citet{mrn77}:  $m = \int^{a_{max}}_{a_{min}} \rho \ N(a) \ V(a) \ da$. \\
		$^{a}$ - \citet{silv10} \\
		$^{b}$ - \citet{poll94}, Table 1B.  Given the temperature of the Local Bubble ($T > 1000$ K), it is assumed all FeS has been converted to Fe. \\
		$^{c}$ - \citet{mats11} and \citet{sj75} \\
		$^{d}$ - \citet{slav10} \\
		$^{e}$ - \citet{mrn77} \\
		$^{f}$ - \citet{hoppe00} \\
		$^{g}$ - \citet{burns79} and \citet{gust94} \\
		}
	}
\end{sidewaystable}
\end{center}

In order to determine the isotope fraction that condenses into dust, we use recent observations of SN 1987A.  {\it Herschel} observations in the far infrared and sub-millimeter wavelengths were modeled in \citet{mats11} with different elemental abundances and dust compositions.  In both models studied, the synthesized Fe mass was nearly identical to the Fe condensed into dust.  This suggests that after less than 30 yrs (much less time than required to travel to a nearby solar system), practically all of the iron from the SN is in the form of dust.  Furthermore, when comparing condensation temperatures \citep{sj75}, one finds that $T_{C, {\rm Al}} = 1800 \ {\rm K} > T_{C, {\rm Fe}} = 1500 \ {\rm K}$, suggesting Al would condense into dust at the same time as Fe, if not sooner.  Based on this reasoning, we assume 100\% of Al and Fe condenses into dust for both SNe and SAGBs.

While refractory elements seem to condense rapidly after ejection, only the dust that survives transport to the Solar System will reach the Earth.  Dust leaving from SAGBs will be subject to shocks from neighboring star systems as well as sputtering from radiation and collisions with other dust grains.  However, for the purposes of this paper, we will assume these affects are negligible compared to other filtering effects examined.  Therefore, we will assume all of the dust from SAGBs are able to pass from the SAGB to the Solar System \citep[a more in-depth discussion of interstellar effects on dust grains is discussed in][]{murr04}.  

Conversely, SN remnants are likely to be much harsher environments for dust, leading to predictions of very small survival probabilities for some dust species and thus for some radioisotopes.  Dust formed from ejecta in a newborn SN remnant will encounter a reverse shock as the remnant transitions from the free expansion to the Sedov/adiabatic phase.  The reverse shock propagates from the outer edge of the remnant back to the source and is generally stronger than the interface between the outer edge of the remnant and general ISM.  The reverse shock causes large-scale sputtering/destruction of grains resulting in gas phase emission from previously refractory elements \citep[e.g., see emission from Cassiopeia A, ][]{rho08}.  \citet{silv10,silv12} studied this interaction and found grains $\lesssim 0.1 \ \mu$m to be most affected; $\sim 1\%$ of Al$_{2}$O$_{3}$ (corundum), $\sim 50\%$ of FeS (troilite), and $\sim 100\%$ of Fe (metallic iron) previously condensed into dust survives.

Once the dust reaches the Solar System, it must pass through the heliosphere to reach the Earth.  \citet{linde00} suggested a cut-off grain size of $0.1-0.2 \ \mu$m for filtering by the heliosphere for grains with speeds of 26 km s$^{-1}$ corresponding to the Sun's motion through the local ISM.  This filtering is less severe for faster dust grains \citep{af11}, but for our larger SN distances we find slower speeds in Table \ref{tab:60Fe_params}.  Since magnetohydrodynamic simulations by \citet{slav10} showed penetration but strong deflection of 0.1 $\mu$m grains, we chose a minimum grain size of 0.2 $\mu$m for entering the Solar System.  For SN dust, this cut-off means that negligible amounts of Al$_{2}$O$_{3}$ and FeS grains will enter the Solar System, while $\sim 10\%$ of Fe grains will be large enough to pass through.  For SAGB dust, we assume the Fe is in elemental Fe and silicate grains (distributed according to \citet{poll94} with FeS assumed to be Fe); Al will be in Al$_{2}$O$_{3}$, but with a larger grain size \citep{hoppe00}.  The Fe size distribution is assumed to be the same as for SN \citep{ster13}, and silicate grains are assumed to follow the \citet{mrn77} distribution ($dN/da \propto a^{-3.5}$) ranging from 0.5-350 nm \citep{wein01}.  This means $\sim 10\%$ of Fe, $\sim 25\%$ of silicates, and $\sim 100\%$ of Al from SAGBs will enter the Solar System.

Lastly, the dust grain must overcome the Sun's radiation pressure once it is in the Solar System.  For this, we consider the parameter $\beta$ \citep{burns79} that characterizes the ratio of the Sun's radiation force, $F_{r}$, to the gravitational force, $F_{g}$:
\beq
	\beta \equiv \frac{F_{r}}{F_{g}}=  C_{\rm r}Q_{\rm pr}\frac{3}{4a\rho_{\rm grain}}
\eeq
where $C_{\rm r}$ is $7.6 \times 10^{-5}$ g cm$^{-2}$ and $Q_{\rm pr}$ is the efficiency of the radiation on the grain \citep[we will assume $Q_{\rm pr} \sim 1$ for the size of grains we are interested in,][]{gust94}.  From \citet{ster13}, only dust grains with $\beta \lesssim 1.3$ will reach Earth's orbit; based on the densities of the minerals considered, if the grain can enter the Solar System, it will be able to reach Earth's orbit.

Combining each of these factors, we find for SNe:  $f_{\rm Fe, SN} \approx 0.01$ and $f_{\rm Al, SN}$ is negligible.  For SAGBs:  $f_{\rm Fe, SAGB} \approx 0.2$ and $f_{\rm Al, SAGB} \approx 1$.  In spite of the number of considerations in determining these quantities, there are still others that could be included, namely a velocity dependence on the filtering by the heliopause.  We would expect dust grains with a sufficiently high velocity (i.e., $v_{\rm grain} > v_{\rm esc,\odot}$) to ignore size filtering limitations, but including these effects will be left for a future work.

\section{Live Radioisotope Data}
\label{sect:data}

\subsection{Terrestrial Measurements of \fe60}
\label{sect:fe60terr}

The primary data value for our analysis is the decay-corrected \fe60 fluence measured by \citet{knie04}
in a deep-ocean Fe-Mn crust.  They found an isotopic ratio of $\fe60/{\rm Fe} = 1.9 \times 10^{-15}$ within the crust; this corresponds to a decay-corrected fluence of ${\cal F}_{{\rm arr}, 60} = (2.9 \pm 1.0) \times 10^{6} \ {\rm atoms \ cm^{-2}}$.  This may be used to determine the distance from the progenitor.  At the time of the original measurements, the half-life of \fe60 was estimated to be 1.49 Myr and that of \be10 was estimated to be 1.51 Myr, resulting in an arrival time, $\tarr = 2.8$ Myr ago.  Since then, the half-lives have been refined:  current best estimates being $\tau_{1/2}$\textsubscript{, \fe60} $ = 2.62$ Myr and $\tau_{1/2}$\textsubscript{, \be10} $ = 1.387$ Myr, and this places the signal arrival at 2.2 Myr ago.  To update the Knie \fe60 fluence, we use ratios to convert the previous results to the updated values that are similar in method to those employed by \citet{be11} and \citet{feige12}:
${\cal F}_{\rm 60, update}/{\cal F}_{\rm 60, previous} = e^{-t_{\rm 60, update}/\tau_{\rm 60, update}}/e^{-t_{\rm 60, previous}/\tau_{\rm 60, previous}}$, which gives the following updated, decay-corrected fluence in the crust:
\beq
\label{eq:N60obs}
	{\cal F}_{\rm arr, 60 (updated)} = (1.41 \pm 0.49) \times 10^{6} \ {\rm atoms \ cm^{-2}}
\eeq

The \fe60 measurement in the crust has been verified by \citet{fit08}, but within this same work,  no comparable \fe60 signal was detected in a sea sediment sample.  
\citet{fit08} suggested several reasons for the non-detection in the sediment, including differences in uptake and divergences in the sediment from the global background.  In addition, we note that the fluence calculation assumes an even distribution of dust over the Earth's surface.  However, the Earth's wind patterns are not uniform, nor is Earth's precessional axis necessarily orthogonal to the progenitor's position.  Consider, for example, a spherical dust grain of radius 0.2 $\mu$m falling at terminal velocity ($\sim 0.1$ m s$^{-1}$) through a 1500 m-thick jet stream flowing horizontally at 100 km hr$^{-1}$ with a density of $4 \times 10^{-4}$ g cm$^{-3}$ (the assumption of falling at terminal velocity should be valid as the Earth's atmosphere will have dissipated most of the dust grain's remaining interstellar kinetic energy).  As the dust grain falls, the pressure from the jet stream will quickly accelerate the grain horizontally to the same velocity and push the grain $\sim 300$ km before it falls out of the jet stream.  In view of the non-uniformity of the jet stream's flow as well as other terrestrial winds, anisotropies in the observed fluences are expected.  Furthermore, one should also consider the source's orientation to the Earth's precessional axis; the Fe-Mn crust used by \citet{knie04} is from 9$^{\circ}$18' N, 146$^{\circ}$03' W ($\sim 1,000$ mi/1,600 km SE of Hawaii), and the sediment used by \citet{fit08} is from 66$^{\circ}$56.5' N, 6$^{\circ}$27.0' W ($\sim 250$ mi/400 km NW of Iceland).  The crust sample's location relative to the equator would make it more likely to receive a signal over a range of arrival angles while the northern hemisphere could be partially shielded from a more southerly progenitor.  Rather than the Fe-Mn crust signal being due to a misinterpretation of the global background, the absence of a sediment signal could be due to the geometry of the source's position.  If this is the case, a sediment sample from the southern hemisphere (e.g., ELT49-53, 38$^{\circ}$58.7' S, 103$^{\circ}$43.1' E and ELT45-21, 37$^{\circ}$51.6' S, 100$^{\circ}$01.7' E used by \citet{feige13}) should have an \fe60 signal.

\subsection{Lunar Measurements of \fe60}

In addition to sea sediment and Fe-Mn crusts, lunar surface (regolith) samples can also be used to search for a nearby progenitor signal.  The lunar surface is not affected by wind or water erosion, but, as pointed out by \citet{feige13}, the sedimentation rate is low (precluding the possibility of time-resolved measurements) and regular impacts by a range of impactors \citep{la77}, continually churn up the regolith, mixing different levels.  Lunar samples would be better suited to providing a ``first hint'' of a signature \citep{feige13}.  Apollo core samples were analyzed by \citet{cook09}, \citet{fimi12}, and \citet{fimi14}; in particular, these authors found both the Apollo 12 sample 12025 and Apollo 15 sample 15008 to have an \fe60 signal above the background.  \citet{nish79} and \citet{nish90} found that the Apollo 12 and 15 cores, respectively, showed little to no large-scale mixing just prior to, during, and/or since the potential arrival of the signal, meaning no large impactor could have ejected part of the regolith thus diluting the signal.  However, as we show in the next section, another issue arises with the dust's arrival at the Moon's surface that can dilute the signature.

\subsection{Lunar Regolith and Dust Grains:  Vaporization}
\label{sect:vapor}

As noted above, there are putative detections of a non-meteoric \fe60 lunar anomaly, which seem to verify the presence of the deep-ocean signal; an amazing confirmation.  However, having argued that the \fe60 will arrive in the form of high-velocity dust grains, we now consider the implications for the deposition onto the lunar surface.

\citet{cin92} made a detailed study of impacts on the lunar regolith and found semi-empirical fits to the volume of vapor produced as a function of impactor size and velocity and of the target composition. It was found that impactor velocity is the dominant factor and that, for $v_{\rm grain} > 100$ km s$^{-1}$, the volume of target material that is vaporized is $\sim 10-100$ times the volume of the impactor itself.  More recently, \citet{crem13} showed that micrometeor impacts can be a non-negligible source of the lunar vapor atmosphere.  They use the \citet{cin92} model and find that “the contribution may be 8\% of the photo-stimulated desorption at the subsolar point, becoming similar in the dawn and dusk regions and dominant on the night side.”  Moreover, \citet{coll14} did laboratory experiments to simulate micrometeorite impacts, studying the neutral gas created as a result.  They find that the number of neutrals produced per unit impactor mass scales as $\sim v_{\rm grain}^{2.4}$, and they conclude that complete vaporization is expected for speeds exceeding 20 km s$^{-1}$.  

With these results in mind, we now consider the conditions surrounding the arrival of the dust signal in question at the lunar surface.  After entering the Solar System, the dust grains continue to the Earth/Moon at essentially the same speed.  However, whereas the Earth's atmosphere slows the arriving dust grains prior to reaching the surface, the Moon's tenuous atmosphere has practically no influence, and the dust grains' velocities are unchanged before arrival at the lunar surface.  The dust grains are estimated to be $\sim 0.2 \ \mu$m in size, and moving at $\sim 20-100$ km s$^{-1}$ depending on the progenitor's distance.  Thus the grains behave as ``micrometeorites'' moving at very high speeds similar to those examined above.

Consider a silicate dust grain ($\rho_{\rm grain} \sim 3.5 \ {\rm g \ cm}^{-3}$) impacting the lunar surface ($\rho_{\rm regolith} \sim 1.6 \ {\rm g \ cm}^{-3}$) at $v_{\rm grain} \sim 20 \ {\rm km \ s^{-1}}$.
The grain arrives with kinetic energy $E_{\rm grain} = m_{\rm grain}v_{\rm grain}^2/2 \sim 0.23 \ \rm ergs$, where
$m = 4\pi \rho_{\rm grain} a^3/3$ is the mass of a spherical grain of radius $a$.
The dust grain will penetrate the regolith to a depth comparable to its diameter and will vaporize some of the surrounding material; 
we will assume $V_{\rm vapor} = 10 V_{\rm grain} \Rightarrow m_{\rm vapor} \approx 4.6 m_{\rm grain}$.  
Given the grain's high speed and shallow penetration, the vaporization will happen quickly (i.e., very little expansion occurs before the entire mass is vaporized).  Moreover, since the initial density of the vaporized regolith is much greater than the density of the Moon's atmosphere, the gas will behave as if it is expanding isentropically into a vacuum.  The grain's kinetic energy will go into vaporizing the grain and regolith and into the thermal and kinetic energy of the resulting gas.  To determine the vaporization energy, the standard enthalpy of formation for the lunar regolith is $\sim 1.5 \times 10^{11} \ {\rm ergs/g} = \left( 3.9 \ {\rm km \ s^{-1}} \right)^{2}$.  This is an approximate value for both of the lunar regolith's main constituents, silica, SiO$_{2}$, and aluminum oxide, Al$_{2}$O$_{3}$, \citep{np73}, and includes both the vaporization and dissociation energies for the molecules.  
Therefore, the total energy consumed in vaporization is 0.1 ergs, leaving 0.13 ergs for the thermal and kinetic energy, $E_{\rm vapor}$, of the gas.

As the gas expands, the thermal portion vanishes asymptotically, with all the energy becoming kinetic.  Thus, after this cooling, the asymptotic expansion speed of the vaporized material is:
\beq
v_{\infty} = \sqrt{\frac{2E_{\rm vapor}}{m_{\rm vapor} + m_{\rm grain}}}
  \approx 6 \ {\rm  km \ s^{-1}} 
\eeq
[For further discussion, see \citet[p. 101-104, 844-846]{zr}].  

The vapor speed is much larger than the lunar escape velocity, $v_{\rm esc, Moon} = 2.4$ km s$^{-1}$.
This suggests that much of the vaporized material, including the dust impactor with its \fe60 material, would escape from the Moon.  This would imply that the Moon has an uptake factor $U_{\rm Moon} \ll 1$.  Thus, we should not be surprised that the lunar results for \fe60 are lower than expected naively from the terrestrial Fe-Mn crust results.  While lunar samples confirm the signal found in the Fe-Mn crust, they are less suitable for determining the fluence given the difficulties in determining $U_{\rm Moon}$.

\subsection{\pu244 Measurements}

Several searches for live \pu244 have been performed, beginning with \citet{wall00} looking at Fe-Mn nodules.  Studies of top layer sea sediment by \citet{paul01}, \citet{paul03}, and \citet{paul07} have shown there is a very low background in \pu244, making \pu244 an excellent candidate to confirm an \fe60 signal from an extra-solar source (presumably from a CCSN).  \citet{wall00, wall04} reported the detection of a single \pu244 atom in the same Fe-Mn crust sample used by \citet{knie99} and \citet{knie04} covering the entire time interval of $1-14$ Myr.  Separately, \citet{rais07} looked in sea sediment for a \pu244 signal, but did not find any evidence for a signal.  It should be noted, however, that the Raisbeck study was using the previous arrival time (2.8 Myr ago) for his search.  The samples were dated using magnetic polarization analysis and did not cover a large enough time interval to include the appropriate dating interval using the new value for the \be10 lifetime \citep{mey94}.

\subsection{\al26 Measurements}

\citet{feige13} reported on searches for \be10 and \al26 using $\sim 3$ kyr time intervals in samples from sea sediments ELT 49-53 and ELT 45-21.  In the case of \al26, the measurements showed only variations consistent with fluctuations around the background level, and found no evidence for an extra-solar signal.  The paper also reported that \mn53 measurements are planned.

\section{Results}
\label{sect:results}

\begin{figure}
	\begin{center}
		\epsfig{file=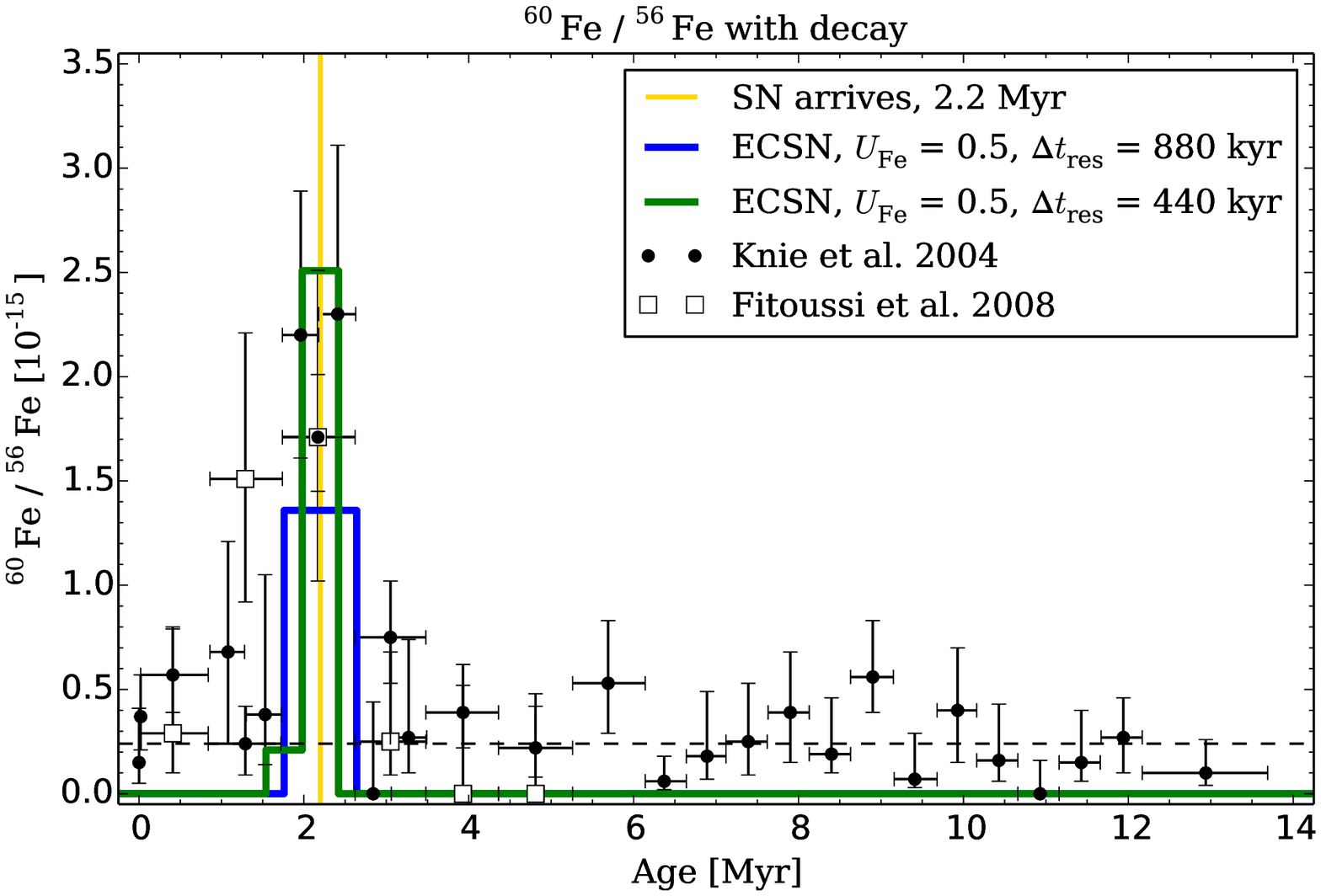,width=\textwidth}
		\caption{\it Evidence from \citet{knie04} and \citet{fit08} for an anomalous peak in the \fe60 isotope fraction $\sim 2.2$ Myr ago, compared with simulations of a possible signal from a SN explosion.  We plot the results using ECSN yields; other progenitors yield similar results.
\label{fig:60fe}
}
	\end{center}
\end{figure}

In Figure \ref{fig:60fe}
we compare our model predictions with the \fe60 data of \citet{knie04} and \citet{fit08}, showing that the model matches the results within the uncertainties for a SN or SAGB occurring 2.2 Myr ago.  We note that the sampling was continuous through the entire data range, and straddled the signal arrival.  In addition, the value for the 880-kyr time resolution was less than the 440-kyr sample, as expected due to the additional stable Fe in the wider sample.

\begin{figure}
	\begin{center}
		\epsfig{file=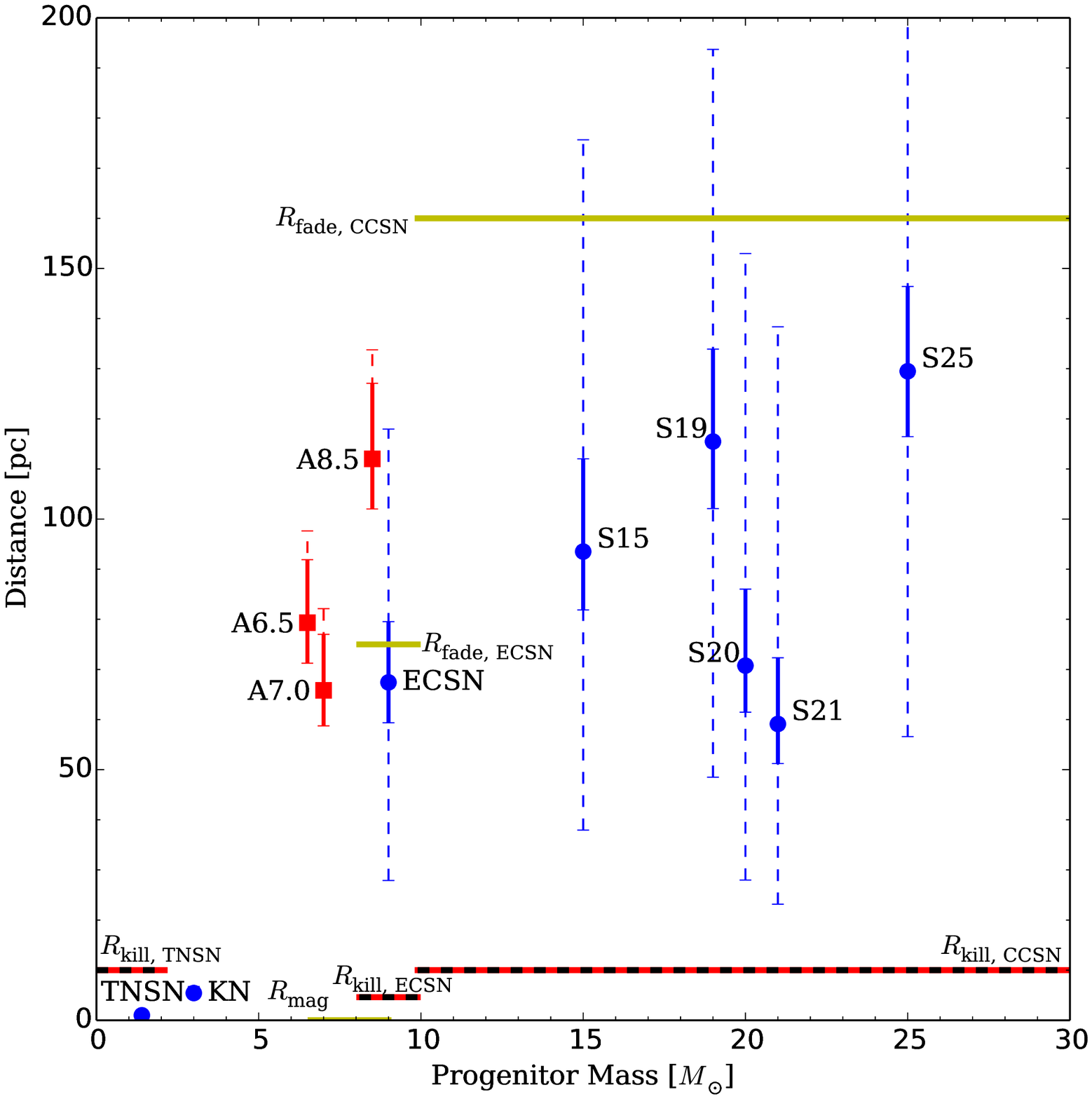,width=\textwidth}
		\caption{\it Estimated distances for possible progenitors, for $U_{\rm Fe} = 0.5$.  SN candidates are circles and SAGB candidates are squares.  The solid error bars represent uncertainty in the fluence measurement \citep{knie04}.  The dashed error bars represent additional uncertainty in \fe60 yields due to nuclear reaction rates in SNe \citep{tur2010} and a delayed super-wind phase in SAGBs \citep{doh13}.  Of particular note are the TNSN/Type Ia SN and the KN/NS-NS merger models, which are too close to have produced the detected \fe60 signal.
		\label{fig:dist}
		}
	\end{center}
\end{figure}

Using the decay-corrected \citet{knie04} fluence of \fe60 (\S \ref{sect:fe60terr}), and \fe60 yields from various source candidates (\S \ref{sect:progenitors}), we have
solved Equation (\ref{eq:Ntotobs}) for the distance to the source.  
Distances and other parameters for some of the possible sources appear
in Table \ref{tab:60Fe_params} and Figure \ref{fig:dist}. 
We see that, for sources at distances $\sim 100$ pc that are typical of our subsequent estimated distances, the en route time and the signal width are ${\cal O}$(Myr), so it is possible that the signal could be time-resolved in future measurements, and thus it is of interest to model the signal shape.

\begin{center}
\begin{table}[*h]
\caption{Predicted Parameters for Possible \fe60 Signal Sources
\label{tab:60Fe_params}
} 
	\begin{tabularx}{\textwidth}{ |X C C C C| }
\hline \hline
			Progenitor			&	Distance to Source, $D$, pc	&	Time en route, $\tsn$, Myr	&	Signal width, $\dtsignal$, kyr	&	Arrival speed, $v_{\rm arr}$, km s$^{-1}$	\\ \hline
			6.5-$\msol$ SAGB	&	79$_{-8}^{+13}$		&	2.8		&	100		&	25	\\ \hline
			7.0-$\msol$ SAGB	&	66$_{-7}^{+11}$		&	2.3		&	100		&	26	\\ \hline
			7.5-$\msol$ SAGB	&	73$_{-8}^{+12}$		&	2.6		&	100		&	25	\\ \hline
			8.0-$\msol$ SAGB	&	97$_{-9}^{+14}$		&	3.5		&	100		&	24	\\ \hline
			8.5-$\msol$ SAGB	&	110$_{-10}^{+15}$	&	4.2		&	100		&	23	\\ \hline
			9.0-$\msol$ SAGB	&	110$_{-10}^{+15}$	&	4.1		&	100		&	23	\\ \hline
			15-$\msol$ CCSN		&	94$_{-12}^{+19}$	&	0.44	&	250		&	84	\\ \hline
			19-$\msol$ CCSN		&	120$_{-13}^{+18}$	&	0.74	&	430		&	61	\\ \hline
			20-$\msol$ CCSN		&	71$_{-9}^{+15}$		&	0.22	&	130		&	130	\\ \hline
			21-$\msol$ CCSN		&	59$_{-8}^{+13}$		&	0.14	&	80		&	170	\\ \hline
			25-$\msol$ CCSN		&	130$_{-13}^{+17}$	&	0.98	&	570		&	52	\\ \hline
			8-10-$\msol$ ECSN	&	67$_{-8}^{+12}$		&	0.61	&	351		&	43	\\ \hline
	\end{tabularx}
	{\it Errors are only for variances in the \citet{knie04} decay-corrected fluence value and do not include variations in nuclear reaction rates (SNe) or delayed super-wind phase (SAGBs).  These parameters are calculated with the Fe uptake factor, $U_{\rm Fe} = 0.5$.}
\end{table}
\end{center}

\subsection{Core-Collapse and Electron-Capture Supernovae}

Figure \ref{fig:dist} shows the calculated distances for our examined CCSNe and ECSN; they range from $\sim 60-130$ pc.  All CCSNe from our set lie outside of the kill distance and within the fadeaway distance for both their average fluence values and errors.  Similarly, the ECSN lies outside the kill distance and within the fadeaway distance (the ECSN kill and fadeaway distances are shorter due to its lower explosive energy).  The ECSN upper error is outside the fadeaway distance, but because SN dust can still travel great distances after decoupling, this is not an absolute limitation.  Based on these distances, either a CCSN or an ECSN could have produced the measured \fe60 signal.  

\subsection{Thermonuclear Supernovae}

TNSN produce so little \fe60 that it would require a TNSN to have been at a distance of $\sim 0.6$ pc in order to produce the signal measured by \citet{knie04}.  This is an implausibly short distance, and any uncertainty in the fluence measurement would not change this determination.  At that range, the TNSN would have killed nearly all life on Earth, so we can exclude a TNSN as the source of the \fe60 signal \citep[in this case, the descreening kill distance for a TNSN is $\sim$10 pc and the ionizing radiation kill distance from 10$^{48}$ ergs of $\gamma$-rays is $\sim$20 pc,][]{smith04}.  Adopting the largest yield ($M_{\rm ej,}\textsubscript{\fe60}\sim 10^{-7} \ \msol$) from \citet{seit13} extends the distance to $\sim 6$ pc, which is still inside the kill radius and does not change this conclusion.

\subsection{Kilonovae}
\label{sect:KNyields}

Our calculations give a possible KN distance of $\sim 5$ pc.  
Of the little that is known observationally or even theoretically
about KNe, we are unaware of any estimates of their 
ionizing radiation output.  In addition, the strength and shape of the shock from ejected material is highly dependent on the orientation of the merger.
Thus, we are unable to estimate the 
corresponding kill distance either by direct exposure or descreening.
The ejecta from KNe are certainly energetic \citep[explosive velocities $\sim 0.3c$,][]{gor11}, and one might imagine decompressing neutron star matter initially emitting in the UV or at shorter wavelengths.
However, the observed radiation for the KN candidate
associated with GRB 130603B 
is very red at times $\ga 8$ hours \citep{berger2013}.
Moreover, while the KN shock and radiation is expected to be much more isotropic
than the GRB,
more study of the geometry of the resulting blast is needed to determine a definitive kill distance like that used for TNSN.  
Consequently, a biohazard argument cannot
rule out a KN explosion as the source of 
the \fe60 anomaly.

However, a much better discriminator for a KN source would be the \pu244/\fe60 ratio.  The single \pu244 atom detected by \citet{wall00,wall04} yields a surface fluence of $3 \times 10^{4}$ atoms cm$^{-2}$ for the period $1-14$ Myr ago.  Looking at the yields from \citet{gor11} again, we can infer the yield for $A = 244$ should be at least on the order of the yield for $A = 60$ (i.e., $(\pu244/\fe60)_{\rm KN} \geq 1$).\footnote{More likely, $A = 244$ yields are 10-100 times larger than $A = 60$ yields given the $A \sim 240$ yields and the fact that the fission recycling sources are centered around $A \simeq 280-290$ region, \citet{gor11}.}  Based on this assumption and the surface fluence for \fe60 during the signal passing ($1.41 \times 10^{6} \ {\rm atoms \ cm}^{-2} / 0.5 = 3 \times 10^{6} \ {\rm atoms \ cm}^{-2}$), then:
\[ 
	(\pu244/\fe60)_{\rm measured} \approx 10^{-2} \ll 1 \lesssim (\pu244/\fe60)_{\rm KN \ predicted}
\]
even though \pu244 was measured over 10 times the time period as \fe60 (Note:  this assumes the dust fraction for Pu is the same as Fe, $f_{\rm Pu} \sim f_{\rm Fe}$).

Additionally, KN occur infrequently \citep[$\sim 10$ Myr$^{-1}$ galaxy$^{-1}$,][]{gor11} compared to CCSN \& ECSN ($\sim 30$ kyr$^{-1}$ galaxy$^{-1}$).\footnote{We would like to thank the reviewer for suggesting this addition to the KNe discussion.}  If we approximate the Milky Way as a thin cylinder of radius 10 kpc and thickness 200 pc, the rates ($\Gamma$) of a KN occurring within $\sim 5$ pc or a CCSN/ECSN occurring within $\sim 75$ pc of the Earth are:
\[
	\Gamma_{\rm nearby \ source} = \Gamma_{\rm galaxy} \frac{V_{\rm nearby \ source}}{V_{\rm galaxy}}
\]
\[ 
	\Gamma_{\rm nearby \ KN} = \left( \frac{1}{10^{7}{\rm \ Myr}} \right) \left(  \frac{D}{5 {\rm \ pc}} \right)^3
\]
\[
	\Gamma_{\rm nearby \ SN} = \left( \frac{1}{1{\rm \ Myr}} \right) \left(  \frac{D}{75 {\rm \ pc}} \right)^3
\]

After inverting these quantities, we can expect a nearby SN every $\sim 1$ Myr compared to a nearby KN every $\sim 10^7 {\rm \ Myr} \gg 1/H_0$ (the Hubble time).  This makes a KN an unlikely source for the \fe60 signal.  However, this result should be revisited as specific yields for \fe60 become available and especially if a signal from strong $r$-process isotopes is detected (e.g., \sm146, \hf182, and \pu244).

\subsection{Super-AGB Stars}

Figure \ref{fig:dist} plots three of our six examined SAGBs (all are listed in Table \ref{tab:60Fe_params}).  Their distances range $\sim 60-110$ pc; similar to those of CCSNe.  With errors, all SAGBs lie well within the distance for dust stopping due to drag 
($\sim 90$ kpc) but well outside the magnetic deflection distance ($\sim 1$ pc).  While it is tempting to rule out SAGBs as a source under the assumption that any dust would be quickly deflected, we have decided to not rule out SAGBs \citep[see, e.g.,][]{fris95, cox03, flor04, fris12} since there is uncertainty in the strength, direction, and uniformity of the Local Bubble's magnetic field.  Depending on the nature of the Local Bubble's magnetic field, charged dust particles could travel with very little deflection.  Instead, we will be examining an alternate search in a future work.

\subsection{\al26 Results}

\begin{figure}
	\begin{center}
		\epsfig{file=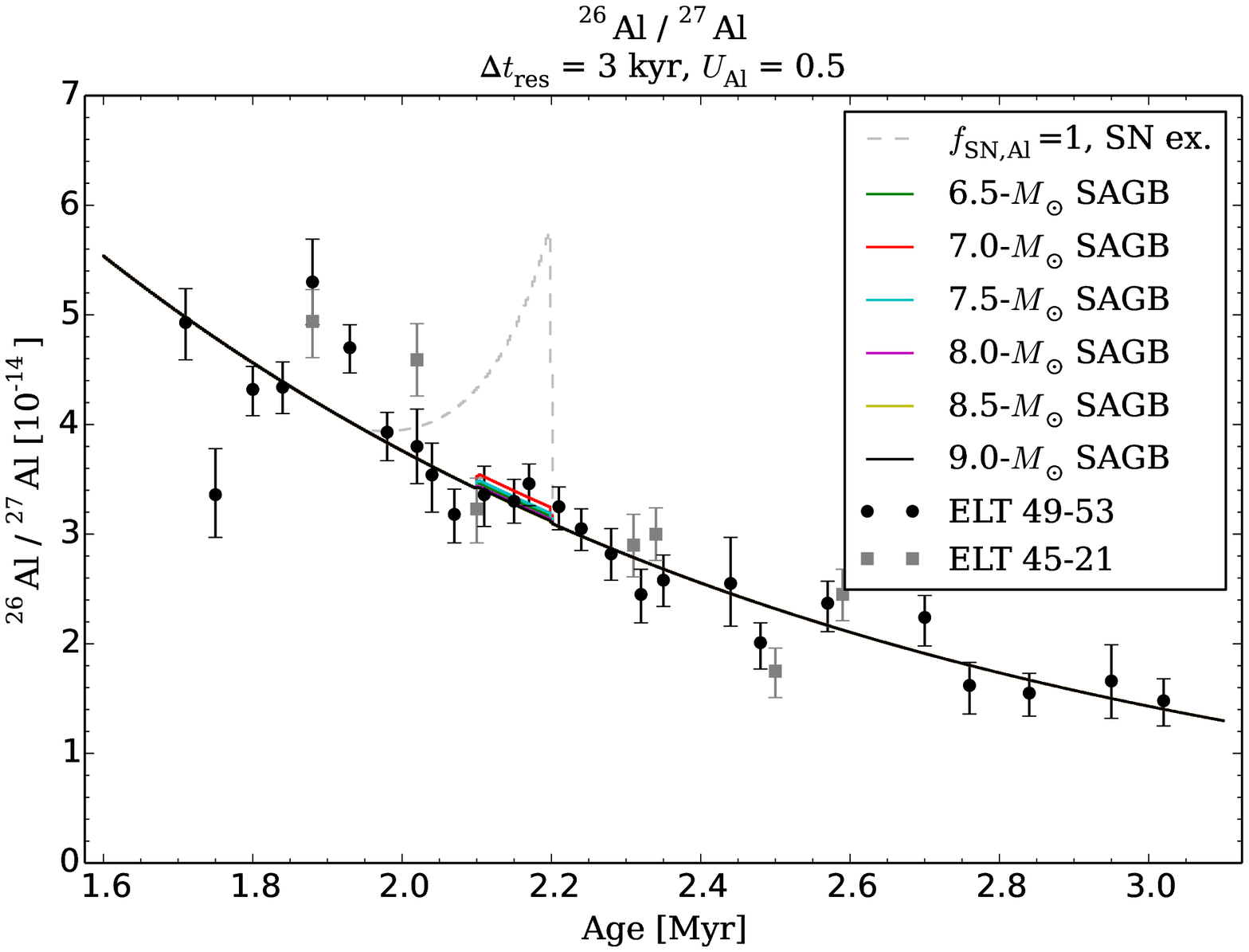,width=\textwidth}
		\caption{\it Model predictions compared with the \al26 AMS data from \citet{feige13}.  Note that the 15-$\msol$ SN is an example only and included to demonstrate the consequence if $f_{\rm SN, Al} = 1$ instead of the $f_{\rm SN, Al} \sim 0$ we expect.
 \label{fig:al26}}
	\end{center}
\end{figure}

In Figure \ref{fig:al26}, we plot \al26 predictions for various progenitors,
and the expected background in the Accelerator Mass Spectrometry (AMS) data from \citet{feige13} using the same 3-kyr sampling intervals used in this experiment.  Because $f_{\rm Al, SN} \approx 0$, we do not expect any signal to be present if the source were a SN.  The calculated signal from a 15-$\msol$ CCSN with $f_{\rm Al, SN} = 1$ is plotted simply as an example if the dust fraction was significantly higher.  Additionally, while we would expect some Al from SAGBs to reach Earth, it would not be visible above the variations in the \al26 background.

\subsection{\mn53 Results}

\begin{figure}
	\begin{center}
		\epsfig{file=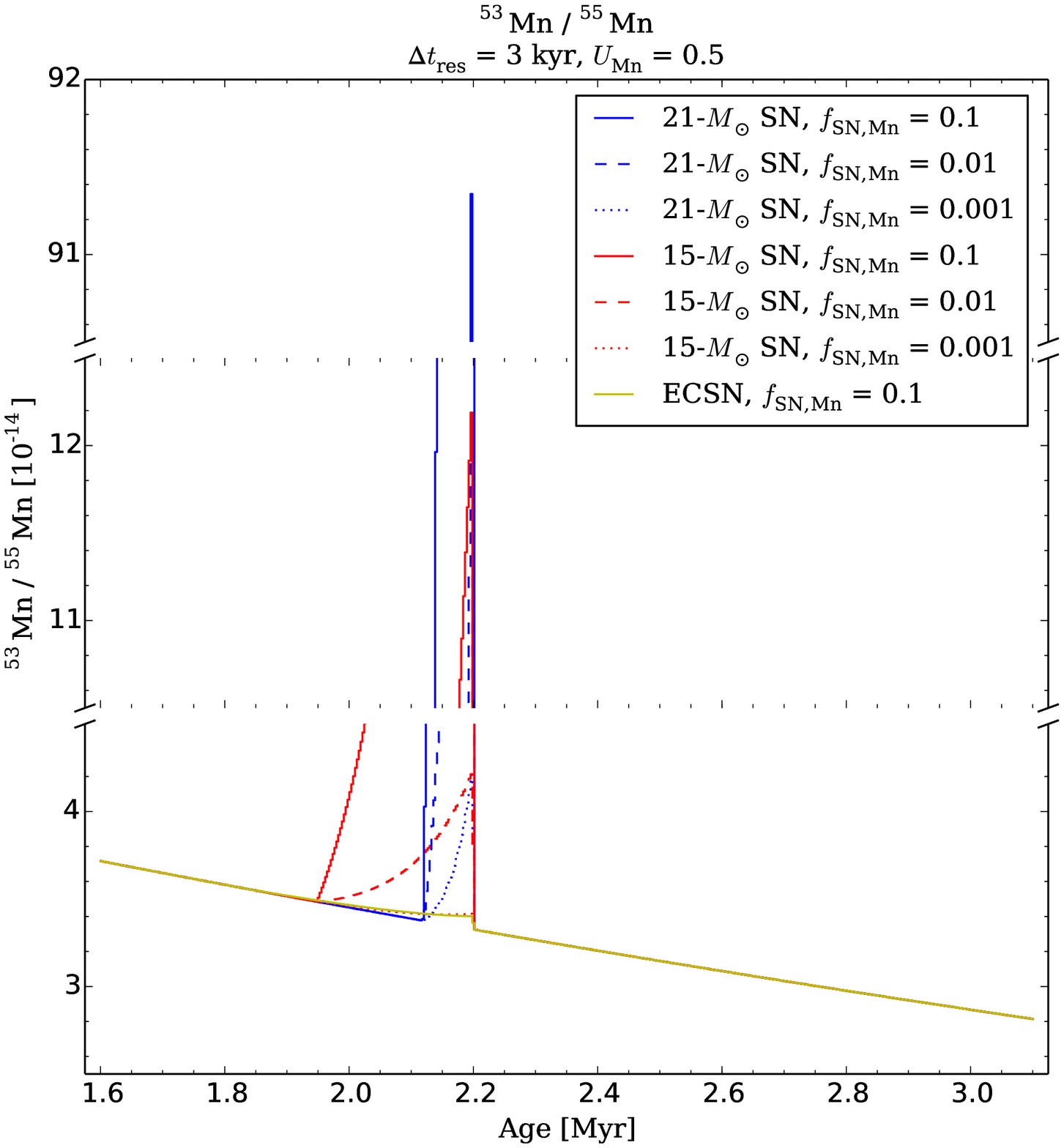,width=\textwidth}
		\caption{\it Model predictions for upcoming \mn53 AMS measurements.  The vertical axis has been broken into three parts in order to show the peak values of each configuration.  Note that the average background \mn53{\rm /}\mn55 level \citep{feige13} is shown ahead of the SN's arrival.  With an AMS sensitivity of $\sim 10^{-15}$ {\rm \mn53/\mn55} \citep{pout10}, progenitors such as a 21-$\msol$ SN should be detectable across a range of $f_{\rm SN, Mn}$ values, whereas an ECSN progenitor should not be detectable.
 \label{fig:mn53}}
	\end{center}
\end{figure}

In anticipation of AMS \mn53 measurements mentioned by \citet{feige13}, in Figure \ref{fig:mn53} we plot predictions for \mn53 based on the distance determined by the \fe60 fluence.  Since the survival and grain size for Mn from a SN has not been described to our knowledge, we plotted a range of SN progenitors (since SAGBs are not expected to produce \mn53) and dust fractions, using the largest possible signal source (21-$\msol$ CCSN), a mid-range source (15-$\msol$ CCSN), and the lowest source (ECSN).  We varied the dust fraction from an order of magnitude above to an order of magnitude below $f_{\rm SN, Fe}$.  As can be seen for the 15- and 21-$\msol$ CCSN with $f_{\rm SN, Mn} \gtrsim f_{\rm SN, Fe}$, a signal should be readily detectable the given the AMS detection threshold of $\sim 10^{-15}$ \mn53/\mn55 \citep{pout10}.  However, for the ECSN and most CCSNe with $f_{\rm SN, Mn} < f_{\rm SN, Fe}$ (even a 21-$\msol$ case could be difficult to detect depending on the fluctuations in the \mn53/\mn55 background), it should be improbable for a SN progenitor to be detected with \mn53.

\section{Discussion and Conclusions}
\label{sect:conclusions}

Since the discovery and confirmation of the terrestrial \fe60 signal by \citet{knie99} and \citet{knie04}, several experiments have tried to find a corroborating signal either in lunar samples or with other isotopes.  To date, none of these experiments provided a definitive signal on the order of that originally reported.  This paper attempts to provide a context for these observations and seek other possible progenitors besides a CCSN whose properties could be consistent with the observations.  We also anticipate future observations with a hope for time-resolved signals.

From our list of candidates, we can rule out a TNSN as it would be too close ($\sim 0.6$ pc) to both create the \fe60 signal and to not kill most life on Earth.  We also rule out a KN as a potential source.  The KN would have been $\sim 5$ pc away from the Earth, and while more study of the geometry of a KN is required to determine a definitive kill distance, the low amount of \pu244 (a strong $r$-process element) detected to date contrasted with the high number of $r$-process elements per merger makes a KN a low probability.  Additionally, KNe/Neutron Star Mergers are very rare, making it unlikely for the Solar System to have passed within 5 pc of one.

Although SAGB stars are outside the magnetic field deflection distance, we have decided to not rule out SAGB stars based solely on this stipulation.  Depending on the strength, direction, and uniformity of the Local Bubble's magnetic field and the charge on the dust grains, it may be possible for a SAGB to have produced the measured \fe60 signal.  Since a SAGB would likely have evolved to the white dwarf stage by now, we plan to investigate this possibility in a future work.

All variations of CCSNe and ECSN remain possible sources.  Of these, ECSN would be the most likely, firstly, because they arise from the lowest-mass and thus most common core-collapse progenitors.
Additionally, \citet{fuchs06} listed members of the Sco-Cen association and their masses (using their listed magnitudes and mass-to-magnitude relation) which included the range:
$M_{\rm Sco-Cen} = 2.5-8.2 \ \msol$, compared to
$M_{\rm ECSN} = 8-10 \ \msol$.  Since more massive stars
evolve faster than lower-mass stars, it is reasonable to expect the
signal progenitor to be near the upper end of the mass range.  Lastly,
the continued lack of a definitive \pu244 signal, in spite of multiple
attempts, is also consistent with the possibility of an ECSN as the
progenitor due to its lack of strong $r$-process products.

Several caveats are important to bear in mind.
Probably most importantly, our ability to 
test different explosion candidates is only as good as
the radioactive yield predictions.  These challenging calculations 
are continually improving, but are subject to significant
uncertainties, including stellar evolution, hydrodynamics, 
and nuclear physics.  
Indeed, two key nuclear cross sections alone
can lead to \fe60 and \al26 yield variations by
factors up to $\sim 10$ \citep{tur2010}.
This alone suggests that all of the CCSN candidates
should be revisited as yields improve.  \citet{seit13} showed that TNSN yields are sensitive to the number of initiation sites and the transition from deflagration to detonation.
Clearly, improved radioisotope yield calculations for any of our explosive sources
could dramatically change the landscape of possible scenarios.
Thus, we implore future nucleosynthesis studies to include
(at least) \fe60 and the other radioisotopes we have discussed here.

Other important uncertainties similarly invite future work.
As we have seen, Fe uptake in Fe-Mn crusts
represent another topic that invites future study.
Uptake has a dramatic impact on our results:
the inferred distance to the explosion
scales as $D \propto U_{i}^{1/2}$.
Additionally, the local interstellar density and magnetic field plays a key role in the propagation of the signal (whether from a SN or SAGB) and in the duration of the time profile of the radioisotope flux.
Finally, as we have seen, the observability of different radioisotopes is highly
sensitive on the formation and survival of supernova
dust of different compositions and sizes.
We have relied on theoretical calculations \citep{silv10,silv12}
which imply, among other things, that \al26 is unlikely to be observable terrestrially
despite its SN abundance comparable to \fe60.  Further such theoretical studies relevant to other radiosotopes,  
and observational corroboration, are critically needed.

To confirm the origin of the \fe60 signal and pin down its source
will require measurements of \fe60 at other sites, other sources (e.g., magnetosomes in addition to crust and sediment), and other radioisotopes.
Lunar regolith measurements provide unique confirmation of the terrestrial \fe60 signal.
However, we find absolute measurements will be difficult because high-velocity dust vaporizes on the lunar surface 
and much of the incident material will then escape the Moon.
This said, we eagerly await detailed presentation of lunar measurements hinted at by \citet{fimi14}.  We are also looking forward to \mn53 measurements as mentioned by \citet{feige13} and \pu244 measurements by \citet[see Wallner, et al. 2014, citation therein]{pir14} that will be helpful in discriminating between the remaining possible progenitors.

Looking ahead to further measurements, the behavior of dust condensation, survival, and filtering will be a key factor in narrowing the remaining pool of possible progenitors.  The dust fraction includes several filtering processes that can all affect the resulting distance calculation ($D \propto f_{i}^{1/2}$).  Studies of dust formation have focused on silicates, iron, and corundum, but formation processes with other elements (especially Mn, Ca, Ti, Zr, Tc, and Pu) could be used to differentiate the remaining possibilities given the varying yield ratios between these elements for each progenitor.  The search for other isotopes is not simply a matter of choosing those with high lifetimes ($\tau_{i} \sim {\cal O}({\rm Myr})$), but also those with low backgrounds, high condensation temperatures (in order to form dust grains), and large grain sizes ($a \gtrsim 0.2 \ \mu{\rm m}$).  Of particular interest would be \ca41 and \mn53.  While perhaps not ideal candidates with regards to background levels, they have long lifetimes and can be condensed at high temperatures ($\gtrsim 1100$ K) into Perovskite (CaTiO$_{3}$), Melilite (Ca$_{2}$Al$_{2}$SiO$_{7}$, Ca$_{2}$MgSi$_{2}$O$_{7}$), and Alabandite (MnS) \citep[see][]{field75}.   Other possible isotopes with long lifetimes such as \zr93, \tc97, \tc99, and \pd107, as well as strong $r$-process elements such as \sm146, \hf182 and \pu244 could be used to constrain CCSNe if more details of their dust condensation are determined, but, regardless, any other candidate isotope would need to form grains large enough to survive escape from its progenitor and enter the Solar System.  It would be a remarkable coincidence if the only isotope that is capable of carrying an extra-solar signal (i.e., \fe60) is the first one examined.

With observations of additional isotopes, it is possible not only to identify a specific event or progenitor, but also to:
(1) provide a better measure of the distance to the source, (2) directly probe individual radioisotope nucleosynthesis, (3) constrain the nearby SN rate, (4) guide astrophysical searches for the SN remains (i.e., pulsars), and (5) model the explosion light curve and to assess the possible damage to the terrestrial biosphere.  Finally, we have seen that the measurement of time-resolved radioisotope profiles provides direct information of the blast passage through the Solar System and an independent measurement of the distance to the progenitor.  The authors are optimistic that new data will make such questions tractable in the near future.

\acknowledgments
Since our first submission, we have been made aware of the superb thesis by Jenny Feige that covers many similar topics and was completed independently of this work \citep{feige10}.  We are pleased to acknowledge the Vera Laboratory including Jenny Feige, as well as the Technische Universit\"{a}t M\"{u}nchen (TUM) Group including Shawn Bishop.  We would also like to thank our reviewer whose thoughtful and thorough comments on the manuscript greatly improved this work.  We are grateful to Thomas Johnson and Craig Lundstrom for illuminating discussions of isotope geology. The work of John Ellis was supported partly by the London Centre for Terauniverse Studies (LCTS), using funding from the European Research Council via the Advanced Investigator Grant 267352.  Brian Fields thanks Stuart Shapiro for his enlightening discussions about the Sedov solution and Friedrich-Karl Thielemann and Ivo Seitenzahl for their informative discussions of supernova radionuclide synthesis.

{}

\newpage

\appendix

\section{List of Variables}

\begin{flushleft}
\underline{Variable - Description [common value or unit of measure]} \\
$a$ - radius of dust grain [$\mu m$] \\
$A$ - atomic number [dimensionless] \\
$\alpha$ - signal width parameter [$\approx 0.577$] \\
$\beta$ - ratio of Sun's radiation force to gravitational force on a particle [dimensionless] \\
$c$ - speed of light [$\sim 3 \times 10^5$ km s$^{-1}$] \\
$c_{s}$ - speed of sound in ISM [km s$^{-1}$] \\
$C_{\rm r}$ - constant from combination of solar flux, gravitational constant, among others [$7.6 \times 10^{-5}$ g cm$^{-2}$] \\
$D$ - distance from progenitor to Earth [pc] \\
$\delta$ - shell thickness of uniform shell for a SN remnant [dimensionless] \\
$E_{\rm CCSN}$ - energy deposited into ejecta by a CCSN [$\sim 10^{51}$ ergs] \\
$E_{\rm ECSN}$ - energy deposited into ejecta by an ECSN [$\sim 10^{50}$ ergs] \\
$E_{\rm KN}$ - energy deposited into ejecta by a KN [$\sim 10^{49}$ ergs] \\
$E_{\rm TNSN}$ - energy deposited into ejecta by a TNSN [$\sim 10^{51}$ ergs] \\
$E_{\rm grain}$ - kinetic energy of dust grain [ergs] \\
$E_{\rm vapor}$ - thermal and kinetic energy in vapor [ergs] \\
$\epsilon$ - shell thickness of saw-tooth shell [dimensionless] \\
$f$ - dust fraction, fraction of isotope that passes from progenitor to Earth [dimensionless] \\
$F_{\rm drag}$ - drag force on dust grain [dyne] \\
$F_{\rm g}$ - force of gravity on dust grain [dyne] \\
$F_{\rm mag}$ - force of magnetic field on dust grain [dyne] \\
$F_{\rm r}$ - force from solar radiation pressure on dust grain [dyne] \\
${\cal F}_{{\rm arr}, i}$ - decay-corrected (or arrival) fluence of an isotope [atoms cm$^{-2}$] \\
${\cal F}_{{\rm interstellar}, i}$ - total fluence of an isotope across spherical signal front [atoms cm$^{-2}$] \\
${\cal F}_{{\rm obs}, i}$ - observed fluence of an isotope [atoms cm$^{-2}$] \\
${\cal F}_{{\rm surface}, i}$ - total fluence of an isotope regardless of uptake [atoms cm$^{-2}$] \\
$\mathbb{F}$ - material flux, fluence per time  [atoms cm$^{-2}$ kyr$^{-1}$] \\
$G(x)$ - density profile function [dimensionless] \\
$G_{0}(s)$ - collisional drag function [dimensionless] \\
$\gamma$ - ratio of specific heats [dimensionless] \\
$\Gamma$ - progenitor rate [Myr$^{-1}$] \\
$H_0$ - Hubble Constant [$\sim 70$ km s$^{-1}$ Mpc$^{-1}$] \\
$i$ - (as subscript) `for a given isotope' (e.g., \fe60, \al26, etc.) \\
$k$ - Boltzmann constant [$1.38 \times 10^{-16}$ erg K$^{-1}$] \\
$m$ - mass [g] \\
$M$ - mass of progenitor [$\msol$] \\
$M_{{\rm ej}, i}$ - total mass of an isotope in the ejecta [$\msol$] \\
$m_{u}$ - atomic mass unit [$\sim 1.66 \times 10^{-24}$ g] \\
$n$ - number density (e.g., of ISM, dust grain, etc.) [cm$^{-3}$] \\
$N$ - number of dust grains [dimensionless] \\
${\cal N}_{i}$ - number of atoms of an isotope [dimensionless] \\
$P_{\rm SW}$ - pressure of solar wind [dyne cm$^{-2}$] \\
$Q_{\rm pr}$ - efficiency of solar radiation on dust grain [dimensionless] \\
$R_{\rm drag}$ - distance at which drag effects are significant on a dust grain [pc] \\
$R_{\rm fade}$ - distance at which a SN shock transitions into a sound wave [pc] \\
$R_{\rm kill}$ - distance from the Sun a SN progenitor can produce a shock that penetrates to Earth's orbit [pc] \\
$R_{\rm mag}$ - distance at which magnetic deflection effects are significant on a dust grain [pc] \\
$\Ra$ - radius of leading edge of SAGB dust shell [pc] \\
$\Rs$ - radius of leading edge of SN remnant [pc] \\
$\rho$ - mass density (e.g., of ISM, dust grain, etc.) [g cm$^{-3}$] \\
$\rho_{0}$ - density in front of shock [g cm$^{-3}$] \\
$\rho_{1}$ - density behind shock [g cm$^{-3}$] \\
$s$ - velocity parameter [dimensionless] \\
$t$ - elapsed time [Myr] \\
$t_{\rm arr}$ - time from today in the past that the leading edge of the signal arrived [Myr] \\
$t_{\rm dep}$ - time from today in the past that the trailing edge of the signal departed [Myr] \\
$\tsn$ - time for isotope to transit from progenitor to Earth [Myr] \\
$T$ - temperature of ISM [K] \\
$T_{C}$ - condensation temperature [K] \\
$\Delta t_{\rm inter}$ - delay between envelope pulses of SAGB [$\sim 100$ yr] \\
$\Delta t_{\rm pulse}$ - duration of envelope pulse of SAGB [$\sim 1$ yr] \\
$\Delta t_{\rm res}$ - time resolution of samples [kyr] \\
$\Delta t_{\rm SAGB}$ - duration of SAGB phase [$\sim 100$ kyr] \\
$\Delta t_{\rm signal}$ - signal width, time for signal to pass Earth and duration of ejecta deposition on Earth [kyr] \\
$\tau_{i}$ - mean lifetime of an isotope, $\tau_{i} = \tau_{1/2, i} / \ln 2$ [Myr] \\
$\tau_{1/2, i}$ - half-life of an isotope [Myr] \\
$U_{i}$ - uptake, fraction of deposited isotope that is incorporated into sampled material [dimensionless] \\
$v$ - speed [km s$^{-1}$] \\
$v_{\rm arr}$ - velocity of a dust grain upon arrival at Earth [km s$^{-1}$] \\
$v_{\rm esc}$ - escape velocity [km s$^{-1}$] \\
$v_{\rm grain}$ - speed of dust grain/impactor [km s$^{-1}$] \\
$v_{\rm grain, 0}$ - initial speed of dust grain [km s$^{-1}$] \\
$v_{\infty}$ - speed at infinity [km s$^{-1}$] \\
$v_{\rm SN}$ - speed of leading edge of the SN remnant [km s$^{-1}$] \\
$V$ - volume [cm$^{3}$] \\
${\cal V}$ - voltage of dust grain [V] \\
$X$ - mass fraction [dimensionless] \\
$\xi_{0}$ - SN proportionality constant [dimensionless] \citep{zr}
\[
\xi_{0} = \left[ \frac{75}{16 \pi} \frac{(\gamma-1)(\gamma+1)^{2}}{(3 \gamma-1)} \right]^{1/5} \stackrel{\gamma=5/3}{\approx} 1.1
\]
$\zeta_{0}$ - SAGB proportionality constant [dimensionless]
\[
\zeta_{0} = \frac{\sqrt{2 \pi \gamma}}{4} \stackrel{\gamma=5/3}{\approx} 0.81
\]
\end{flushleft}

\section{Blast Expansion and Radioisotope Flux Profile}
\label{sect:appendix}

\begin{sidewaysfigure}
	\begin{center}
		\epsfig{file=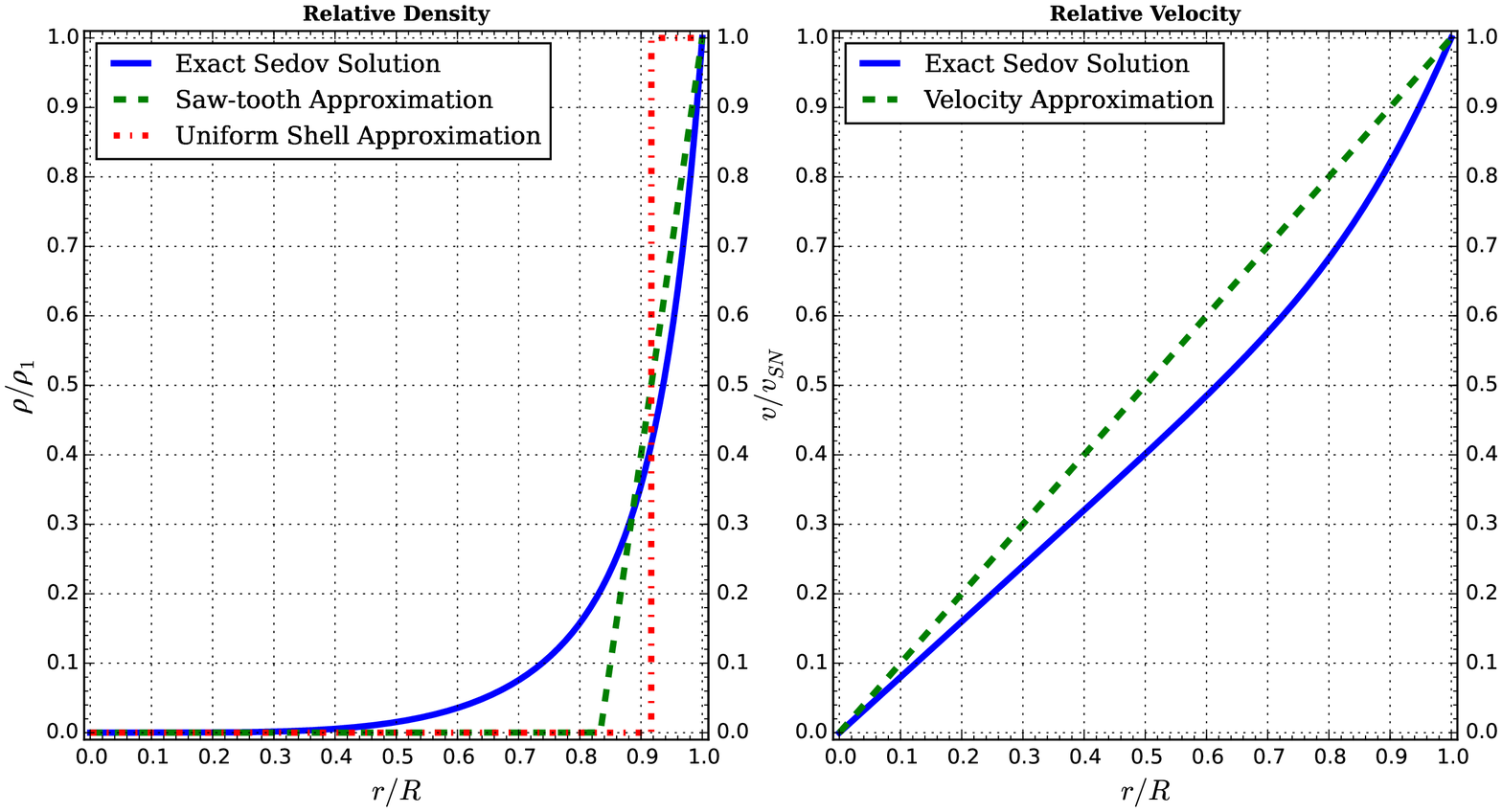,width=\textwidth}
		\caption{\it Comparison of the uniform shell and saw-tooth profiles to exact solution profile, for $\gamma = 5/3 \Rightarrow \delta \approx 0.083$ (uniform shell), $\epsilon \approx 0.17$ (saw-tooth).  In addition, the paper used a ratio to approximate the observed SN crossing velocity.  The right chart compares this approximation to the exact solution profile.
\label{fig:sedov-approx}}
	\end{center}
\end{sidewaysfigure}

We model astrophysical explosions as spherically symmetric, 
and we are interested in distances sufficiently large
that the swept up interstellar mass is much larger than the
ejecta mass.
We treat a blast wave as adiabatic (energy-conserving) and
thus adopt the Sedov-Taylor solution.
The Sedov blast wave evolves in a self-similar manner.  This means in particular that gas properties as a function of radius $r$ maintain the same shape when plotted in terms of the similarity variable:
\[
	x = \frac{r}{\Rs(t)}
\]
where the shock radius at $t$ is given by Equation (\ref{eq:sedov}).
In particular, the density profile is:
\[
	\rho(r,t) = \rho_1 \ G(r/\Rs)
\]
where the density immediately behind the shock is:
\[
	\rho_1 = \frac{\gamma+1}{\gamma-1} \rho_0
\]
and the dimensionless density profile function is thus normalized to $G(1) = 1$.  Note that mass conservation implies that the total mass $M_{swept}=4\pi \rho_0 \Rs^3/3$ swept up in the blast is equal to the total mass in the blast profile:
\[
	M_{total} = 4\pi \int_0^{\Rs} r^2 \ \rho(r,t) \ dr = 4\pi \rho_1 \Rs^3 \int_0^{1} x^2 \ G(x) \ dx
\]
and so setting $M_{total}=M_{swept}$ implies that:
\beq
\label{eq:massint}
	\int_0^{1} x^2 \ G(x) \ dx = \frac{\rho_0}{3\rho_1} = \frac{1}{3} \frac{\gamma-1}{\gamma+1} \stackrel{\gamma=5/3}{\longrightarrow} \frac{1}{12}
\eeq

We consider two approximations to the full Sedov profile.  For a uniform shell approximation, we have $G(x) = 1$ for $x\in [1-\delta,1]$ and zero otherwise, which gives the location of the inner shell radius via:
\[
	\frac{1-(1-\delta)^3}{3} =  \frac{1}{3} \frac{\gamma-1}{\gamma+1}
\]
and thus:
\[
	\delta = 1-\pfrac{2}{\gamma+1}^{1/3} \stackrel{\gamma=5/3}{\longrightarrow} 0.0914
\]
whereas going to first order in $\delta$ we would find $\delta = 1/12$.  
For a ``saw-tooth'' approximation, the blast material is in a
thin shell with a profile that linearly decreases from a maximum behind the shock
to zero at coordinate $x_0 \equiv 1-\epsilon$.
Thus we have $G(x) = Ax + B$, with the constraints that $G(1) = 1$ and $G(x_0) = 0$ at the inner radius, which gives:
\beq
\label{eq:saw}
	G(x) = \frac{x-x_0}{1-x_0}
\eeq
Choosing $\gamma = 5/3$, to first order we find 
$\int_0^1 x^2 G(x)\ dx = \int_{1-\epsilon}^1 x^2 G(x)\ dx \approx \epsilon/2$.
From Equation (\ref{eq:massint}) we find the
dimensionless shell thickness $\epsilon \approx 1/6$, twice the value in the uniform shell.  
As seen in 
Figure \ref{fig:sedov-approx}, 
the saw-tooth density profile more closely matches the exact Sedov density profile compared to the uniform shell profile, making the saw-tooth profile more appropriate for modeling the signal spreading for our SN distances. 

The (radial) velocity profile is:
\[
	v(r,t) = \dot{R}_{\rm s} \ U\pfrac{r}{\Rs}
\]
with $\dot{R}_{\rm s}$ the shock speed, and the dimensionless velocity profile function normalized to $U(1) = 1$.  To a good approximation, the velocity is linear, and we will adopt the approximation $U(x) \approx x$.  This leads to a ``Hubble law'' relation:
\[
	v(r,t) \approx \dot{R}_{\rm s} \ \frac{r}{\Rs}
\]
Figure \ref{fig:sedov-approx}
compares this linear velocity profile
with the exact Sedov solution.  Our approximation is necessary to maintain the self-similarity of the saw-tooth profile, and, while different than the exact solution, should be sufficient for the region we are most interested in ($0.8 \leq r/R \leq 1$).  For a more detailed description of the analytical Sedov solution, see \citet{bk1994}.

Turning to the explosive ejecta,
we note that if a number ${\cal N}_i$ of atoms of species $i$ were distributed with {\em uniform} density at time $t$, then the mean number density in $i$ would be:
\[
	n_{i,0} = \frac{3}{4\pi} \frac{{\cal N}_i}{\Rs^3}
\]

We will assume that, at times of interest, the ejecta is well-mixed into the blast wave, with a constant mass fraction at all radii.  That is, we assume that the ejecta density profile
follows that of the blast itself.  This means that the highest ejecta density is just behind the shock, with a value:
\[
	n_{i,1} = n_i(\Rs) = \frac{\gamma+1}{\gamma-1} \ {n}_{i,0}
\]
and the ejecta density profile is:
\[
	n_{i}(r,t) = n_{i,1} \  G(r/\Rs)
\]

Combining the ejecta density profile with the ``Hubble law''
velocity approximation gives
the global-averaged ejecta flux onto the surface of the Earth (i.e.,
1/4 the interstellar flux, not including radioactive decay), evaluated at distance $r=D$:
\beq
\mathbb{F}_i(D,t) = \frac{1}{4} n_i(D,t) \ v(D,t) = {\cal F}_{1} \ \pfrac{D}{\Rs}^3 \ G(D/\Rs) \ \frac{\dot{R}_{\rm s}}{\Rs}
\eeq
with the time-independent prefactor:
\beq
	{\cal F}_1 = \frac{3}{16\pi} \frac{\gamma+1}{\gamma-1} \frac{M_{{\rm ej,}i}/m_i}{D^2}
\eeq
We see here explicitly that a time-resolved flux
directly encodes the blast density profile 
and thus probes the propagation of the radioisotope ejecta
from explosion to Earth.

Using our saw-tooth approximation for the blast density 
profile $G$ (Equation \ref{eq:saw}),
and using the Sedov result $\dot{R}_{\rm s}/\Rs = 2/5t$, 
we find a flux profile in time of:
\beq
\mathbb{F}_i(D,t) =  \frac{2{\cal F}_{1}}{5t} \ \pfrac{D}{\Rs}^3 \ 
 \frac{D/\Rs(t)-1+\epsilon}{\epsilon}
\eeq

We note that the leading edge of the blast 
from an event at distance $D$ arrives
at a time $t_{i}$ given by $D = \Rs(\tinit)$.
Thus we can recast $D/\Rs(t) = (t/\tinit)^{-2/5}$ in terms of
the initial arrival time.
The trailing edge of the shell arrives at 
time $\tfin$ given by $D = (1-\epsilon) \Rs(\tfin)$.
Thus we have
$\tfin=\tinit/(1-\epsilon)^{5/2}$.
This means that we can write $\epsilon = 1-(\tinit/\tfin)^{2/5}$,
and we can express the global-averaged flux time profile as:
\beqar
\mathbb{F}_i(D,t) =  \pfrac{t}{\tinit}^{-11/5} \ 
 \left[ 
   \frac{(t/\tinit)^{-2/5}-(\tfin/\tinit)^{-2/5}}{1-(\tfin/\tinit)^{-2/5}}
 \right] 
 \ \mathbb{F}_i(D,\tinit)
\eeqar
This is the sum of two power laws in $t$, leading to
a steep cusp at early times $t \rightarrow \tinit$ that flattens 
at late times $t \rightarrow \tfin$.



\end{document}